\documentclass[acmsmall,screen]{acmart}
\AtBeginDocument{%
  }

\setcopyright{acmlicensed}
\copyrightyear{2026}
\acmYear{2026}
\acmDOI{XXXXXXX.XXXXXXX}
\acmConference[Conference 'XX]{conference title}{June 03--05,
  2018}{Woodstock, NY}
\acmISBN{978-1-4503-XXXX-X/2026/10}

\usepackage{algorithm}
\usepackage{algpseudocode}
\usepackage[dvipsnames]{xcolor}
\usepackage{longtable}
\usepackage{booktabs}
\usepackage{tikz}
\usetikzlibrary{arrows.meta,positioning}
\usepackage[normalem]{ulem}
\usepackage{graphicx}
\usepackage{wrapfig}
\usepackage{makecell}
\usepackage{multirow}%
\usepackage{multicol}
\usepackage[table]{xcolor}
\usepackage{xcolor}
\usepackage{tcolorbox}
\usepackage[skip=0cm]{caption}
\usepackage{lipsum}
\usepackage{hyperref}

\begin{document}

\title{Solving Subgraph Extraction Problems Using $\Delta$Search}


\author{Rebin Silva Valan Arasu}
\affiliation{%
  \institution{University of California Riverside}
  \city{Riverside}
  \state{California}
  \country{USA}}
\email{rebinsilva.valanarasu@email.ucr.edu}
\orcid{0000-0002-0573-4893}

\author{Rajiv Gupta}
\affiliation{%
  \institution{University of California Riverside}
  \city{Riverside}
  \state{California}
  \country{USA}}
\email{rajivg@ucr.edu}
\orcid{0000-0002-9348-3974}


\renewcommand{\shortauthors}{R.S. Valan Arasu et al.}

\begin{abstract}
    Many NP-hard graph problems can be modeled as optimal subgraph extraction problems with feasibility constraints. From Network Design to Facility Location, from Robotics to Graph Drawing, the subgraph extraction pattern emerges across diverse domains. Despite this commonality, these problems are typically solved with domain-specific heuristics. Usually, these problems balance competing objectives such as maximizing coverage or minimizing cost while satisfying structural constraints such as connectivity, planarity and reachability. In this work, we introduce $\Delta$Search, a general and fast heuristic framework that exploits the insight of Reward-Penalty optimization for solving a large class of subgraph extraction problems. The framework is easy to use as it only requires feasibility constraints and optimality criteria to be provided by the user to express the subgraph extraction problem. We also show how exact methods can be augmented with $\Delta$Search to improve their performance by aggressive pruning of the search space. We evaluate our framework on monotone graph problems such as Maximum Planar Subgraph (MPS) and Minimum Connected Dominating Set, Weighted Monotone problems such as Maximum Weighted Independent Set and Minimum Weighted Steiner Tree, and non-monotone graph problems such as Prize Collecting Vertex Cover (PCVC) and Uncapacitated Facility Location Problem (UFLP). Our results show that $\Delta$Search matches or surpasses state of the art heuristics for MPS, UFLP and PCVC problems with similar runtime. For the remaining problems, $\Delta$Search achieves approximately $89$\% of the solution quality of the state-of-the-art algorithms without any problem-specific tuning.
\end{abstract}



\received{17 March 2026}

\maketitle

\section{Introduction}



NP-hard \textit{subgraph extraction} problems are abundant in computer science with applications across multiple domains such as Bioinformatics~\cite{amir1997maximum, yu2006predicting}, Network Design~\cite{dai2004extended,yigit2021performance, bhattarai2022steinerlog}, Robotics~\cite{hauser2014minimum}, Logistics~\cite{xu2020approximation, chudak2003improved} and other domains. Thus, there is a need for a general framework that domain experts can use without needing expertise in graph algorithms. A graph framework that only requires what makes a good solution and not how to search for it would be ideal. Therefore, the goal of this paper is to present a novel and efficient way of solving this broad class of subgraph extraction problems with just the problem statement and without any problem-specific tuning or implementation. 

Current graph frameworks are either exact algorithms with exponential worst-case runtime that require minimal user effort~\cite{russell2020modern, almohamad2002linear, trukhanov2013algorithms} or approximate heuristics that are scalable but require extensive problem-specific tuning~\cite{gonzalez2007handbook, grossmann2023finding, sonucc2023adaptive,zhang2023fast, sonucc2021binary}. Numerous heuristics and metaheuristics have been developed to tackle the intractability of graph problems. Heuristics lack generality as they are tailored to individual problems. Although metaheuristics are designed to solve a large class of problems, encoding these problems into a form suitable for these metaheuristics is non-trivial~\cite{gonzalez2007handbook}. The user must perform these complex problem encodings to use these metaheuristics, and thus metaheuristics achieve generality and practical runtime at the cost of increased user effort for each new problem instance. Multiple exact works have also been proposed. Constraint Programming (CP)~\cite{russell2020modern, brailsford1999constraint, dechter2003constraint} and Mixed Integer Linear Programming (MILP)~\cite{almohamad2002linear, aneja1980integer, dias2025minimum} are two general exact exponential frameworks that can solve multiple graph problems including Facility Location, Graph Partitioning and Subgraph selection. But exactness comes at the cost of performance. These exact frameworks can typically scale only to graphs with $100$ to $200$ edges~\cite{chimani2019exact, cimikowski1994branch}. Thus, these frameworks are slow and graph-structure unaware, and they require problem-specific tuning to achieve good performance. 

The framework that is closest to achieving generality is Local Ratio~\cite{bar2004local}, an approximate algorithm that can be applied to multiple graph problems. Local Ratio achieves this by a user-provided approximate algorithm for simpler instances of the problem, which is then repeatedly applied to the given graph to achieve the same approximation ratio as the simpler instance. Although this approach achieves generality, it requires the presence of a problem-specific approximate algorithm for simpler instances and user effort to generate this approximate algorithm.

From the above works, we can conclude that although ease of use and generality are known to be important for subgraph extraction frameworks, most frameworks sacrifice either one or both for computational efficiency. This relentless pursuit of computational runtime has also resulted in most of the frameworks being heavily tailored to the problem instances, limiting their generality.

In this paper, we introduce $\Delta$Search, a subgraph extraction framework that only takes graph feasibility constraints and objective functions and produces good heuristic solutions for a large class of graph problems with solution quality competitive with tailored heuristics. Since both of these functions are usually part of the problem statement itself, the burden on the user is greatly reduced. Through its simpler unified interface, $\Delta$Search sidesteps the issues faced by heuristics and metaheuristics. Although general approximation ratios cannot be given for all the problems solvable by $\Delta$Search, it achieves approximations comparable to the greedy algorithm, since both the algorithms will always find maximal or minimal solutions. For example, $\Delta$Search inherits the $1/3$-approximation of the greedy algorithm~\cite{chimani2009non} for the Maximum Planar Subgraph problem~\cite{junger1994polyhedral}. 

Most optimal subgraph extraction problems require both balancing competing objectives such as maximizing coverage or minimizing costs and satisfying structural constraints such as connectivity, planarity, and reachability. Thus, the central idea of $\Delta$Search is that many subgraph extraction problems can be expressed as finding a subset of graph elements such as vertices or edges that maximizes the difference between reward and penalty functions. We refer to this Reward-Penalty decomposition, which we introduce, as the \emph{Difference of Monotone} (DoM) formulation. This formulation is expressive enough to subsume existing frameworks such as Hereditary graph problems~\cite{borowiecki1997survey}, Ancestral graph problems~\cite{natanzon2001complexity}  and also cover other graph problems (Prize Collecting Arc Routing problem~\cite{araoz2009clustered}, Uncapacitated Facility Location Problem~\cite{chudak2003improved}).

Formulating graph problems as \emph{Difference of Monotone} (DoM) subgraph optimization problems is compelling for several reasons. First, it aligns closely with real world problem objectives such as cohesion, coverage, and structural feasibility, which typically increases or decreases monotonically as graph elements are added. Second, the Reward-Penalty idea of DoM formulation is simpler and intuitive than more complicated submodularity or convexity assumptions. Third, the DoM formulation covers a wide range of problems as can be seen from Table~\ref{table:problems}, enabling the development of general-purpose frameworks and solvers that apply broadly across graph domains.

\begin{figure}[!t]
  \centering
  \includegraphics[width=0.65\linewidth]{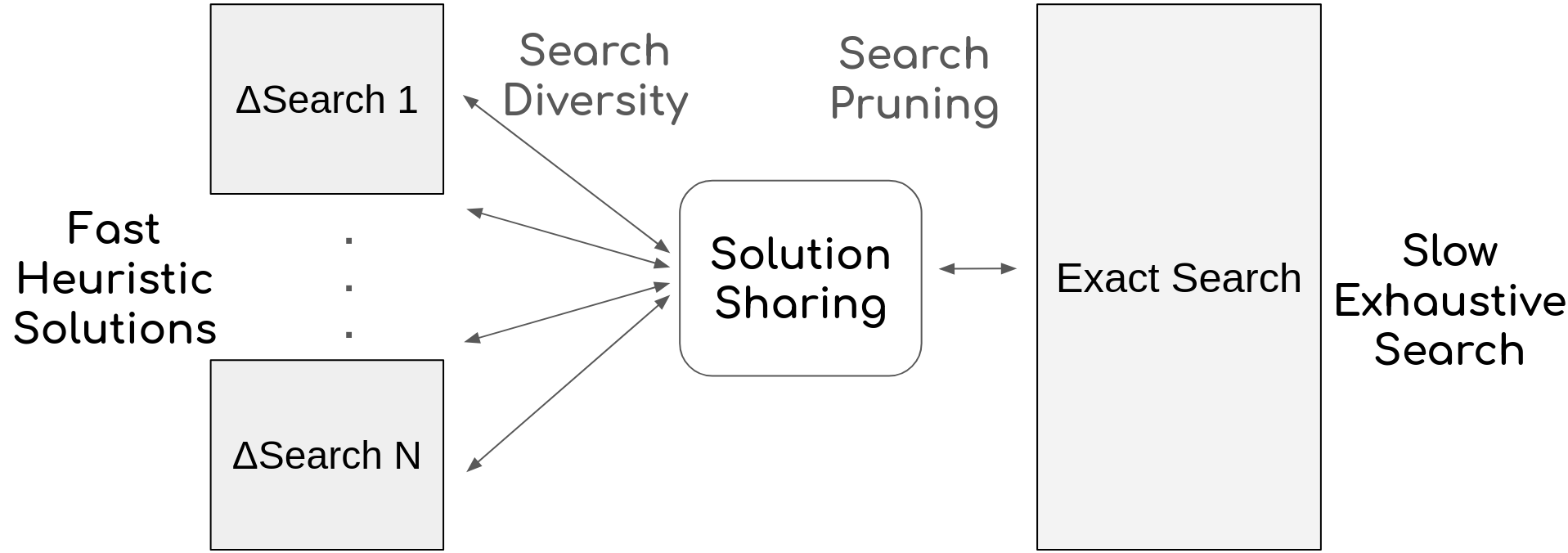}
  \vspace{0.05in}
  \caption{Exact algorithm augmented with the $\Delta$Search Framework.}
  \label{fig:overview}
  \vspace{-0.2in}
\end{figure}

$\Delta$Search performs a binary search-like exploration in the DoM search space. This simplicity in the core algorithm is what allows $\Delta$Search to be both efficient and general. Just as Fuzz Testing~\cite{fuzzingbook2024} generates many simple test cases rather than computing highly complex ones, we realized that running a simpler algorithm like $\Delta$Search multiple times usually outperforms designing a single highly engineered algorithm. Thus, we designed $\Delta$Search which runs the scoring (optimality) function at most quadratic times to the number of graph elements. 

To further demonstrate the usefulness of $\Delta$Search, we show how it can augment exact methods, such as the \emph{Russian Doll Search}~\cite{trukhanov2013algorithms} method, to help prune the exponential search space (see Figure~\ref{fig:overview}). The augmented exact method works by mutually sharing solutions between the exact and heuristic $\Delta$Search, thereby pruning the search in each and accelerating them both. By sharing the solutions, $\Delta$Search is directed toward regions not yet explored by the exact algorithm, and the exact algorithm prunes its explorations based on solutions found by the faster $\Delta$Search. Although exponential time algorithms such as Branch and Bound~\cite{morrison2016branch} can be applied to multiple problems, by combining these exact algorithms with the fast $\Delta$Search heuristic, we accelerate their search by aggressive pruning. While problem-specific pruning heuristics exist, to the best of our knowledge, this work is the first to accelerate an exact algorithm using a general heuristic.
 
Even though problems like the shortest path problem can be converted into a hereditary problem and then expressed in our framework, the benefit of using our framework for the shortest path is limited since our framework cannot compete with exact linear time algorithms tailored for shortest path. 
The $\Delta$Search framework is best suited for NP-hard graph problems whose objective function can be expressed as a DoM function. However, expressing the objective function as a DoM function may not always be possible. If the objective function cannot be expressed as a DoM function, but can still be approximated by one, $\Delta$Search can still be employed.

In summary, this paper makes the following contributions:

\begin{enumerate}
    \item We introduce a new formulation for graph problems called \emph{Difference of Monotone} (DoM) that encompasses a large class of graph problems including ancestral (e.g., Minimum Connected Dominating Set~\cite{dai2004extended}, Minimum Steiner Tree~\cite{kou1981fast}), hereditary (e.g., Maximum Planar Subgraph Problem~\cite{junger1994polyhedral}, Maximum Weighted Independent set~\cite{tarjan1977finding}), and some other non-monotone graph problems (e.g., Prize Collecting Vertex Cover problem~\cite{karp2009reducibility}, Uncapacitated Facility Location Problem~\cite{chudak2003improved}). This new formulation generalizes the monotone graph problem class such as Hereditary and Ancestral graph problem classes while preserving the opportunities for efficient search space pruning explored in previous works~\cite{trukhanov2013algorithms}. 

    \item We introduce $\Delta$Search, which can find approximate solutions for DoM subgraph optimization problems with $O(n^2)$ calls to the scoring function. $\Delta$Search also provides a simple interface where the user only needs to supply the scoring functions to the framework. Thus, the user can rapidly find solutions to wide variety of graph problems easily with only at most $O(n^2)$ calls to the scoring functions.

    \item We evaluate our framework on six subgraph selection problems: Maximum Planar Subgraph (MPS), Minimum Weighted Steiner Tree (MST), Maximum Weighted Independent Set (MWIS), Minimum Connected Dominating Set (MCDS), Prize Collecting Vertex Cover (PCVC) and  Uncapacitated Facility Location Problem (UFLP). Experimental evaluations showed that $\Delta$Search matches or surpasses existing algorithms for MPS, UFLP and PCVC problems and also achieves approximately $89$\% of the best known solution quality for other problems.

    \item We demonstrate the efficiency of $\Delta$Search as a heuristic bound that accelerates exact frameworks. We integrate $\Delta$Search into an exact framework to reduce the search space of the exact algorithm by pruning search directions leading to inferior solutions. Our experiments show that the $\Delta$Search augmented exact algorithm achieves a $2.639 \times$ speedup over the exact algorithm for the Maximum Planar Subgraph problem.
    

\end{enumerate}

The remainder of this paper is organised as follows. Section~\ref{sec:dom} presents the DoM formulation and its applications. Section~\ref{sec:deltasearch} describes the $\Delta$Search algorithm and its complexity. Section~\ref{sec:interface} presents our system design and explains how the interface works. Section~\ref{sec:experiments} reports the experimental results of $\Delta$Search across six graph problems and Section~\ref{sec:deltaexact} explains how $\Delta$Search can be combined with exact algorithms to accelerate the exact search. Section~\ref{sec:relatedwork} surveys prior work on general graph optimization frameworks. Finally, Section~\ref{sec:conclusion} concludes the paper.

\vspace{-0.1in}
\section{Difference of Monotone (DoM) Framework}
\label{sec:dom}

Most real world subgraph extraction problems involve optimizing an objective function while minimizing cost. Examples of such problems include facility location problems where the goal is to maximize service coverage while minimizing facility setup and operational costs, or vehicle routing problems where the goal is to maximize the number of deliveries while minimizing time or fuel spent. These problems can be naturally modeled as the difference between a monotonically increasing reward function and a penalty function. The monotonicity is important as it captures the intuition that adding elements to a subgraph should not decrease the reward or penalty, and it also enables efficient search strategies.

In simpler terms, the Difference of Monotone (DoM) optimization function is a function that can be expressed as the difference between a reward and a penalty function i.e., a function that can be expressed as $R(SG) - P(SG)$ where the reward function $R$ and the penalty function $P$ are monotonically increasing functions. In mathematical terms, any problem of the following form can be solved using this framework: \textbf{Find $SG \subseteq G$ that maximizes $R(SG) - P(SG)$, where $R$ and $P$ are monotone non-decreasing functions.}

Examples of problems with DoM functions can be seen in Table~\ref{table:problems}. Additionally, Figures~\ref{fig:monotone_graph} and \ref{fig:dom_graph} show the behavior of these problems when their respective graph elements are added in some random order. The plots illustrate that the difference of the reward and penalty usually rises initially with graph element additions but later falls down due to the high penalty. Figure~\ref{fig:dom_graph} also shows that there might be multiple maxima for the difference between reward and penalty. Finally, note that these maxima pertain to this particular order of addition of graph elements; the global maximum might not be present in one of these maxima and a different ordering might be able to produce the global maximum for these problems.

As we shall see in this subsection, the DoM framework naturally generalizes Hereditary~\cite{yannakakis1978node} and Ancestral~\cite{natanzon2001complexity} graph problems that have been studied in prior works. It also covers graph problems such as Prize Collecting Vertex cover (PCVC) and Uncapacitated Facility Location (UFLP) problems that are neither hereditary nor ancestral. Table.~\ref{table:problems} lists multiple graph problems that can be formulated as DoM problems and its subclasses Hereditary and Ancestral problems. Note that the class of Constraint Optimization Problems (COP) and Mixed Integer Linear Programming (MILP) problems is much more general than the DoM framework. Thus, DoM framework achieves a balance between generality and structure for efficient search.

\begin{table}[!t ]
\caption{$CP \supset MILP \supset DoM \supset (Hereditary \cup Ancestral)$ \& $Hereditary \cap Ancestral = \emptyset$. Note that $CP$, $MILP$ and Hereditary are all NP-hard problems.}
\label{table:problems}
\small
\begin{center}
\begin{tabular}{|p{0.25\linewidth}|p{0.6\linewidth}|}
\hline
Class of Problems & Graph Problems\\
\hline
\hline
Constraint Optimization Problems (COP) & Graph coloring~\cite{gualandi2012exact}, Traveling Salesman~\cite{vali2017constraint}, Balanced Graph Partitioning, Subgraph Isomorphism~\cite{zampelli2010solving} \\
\hline
 Mixed Integer Linear Programming (MILP)  & Capacitated Facility Location~\cite{wu2006capacitated}, Minimum Flow Decomposition~\cite{dias2022fast}, Graph Edit Distance~\cite{d2025enhancing}, Crossing Number Problem~\cite{chimani2016ilp}, Minimum Chordal Completion~\cite{bergman2015benders}, Capacitated Minimum Spanning Tree~\cite{gouveia19952n}, Clique Partitoning Problem~\cite{koshimura2022concise} \\
\hline
Difference of Monotone (DoM) & Uncapacitated Facility Location~\cite{chudak2003improved}, Prize Collecting Steiner Tree~\cite{bhattarai2022steinerlog}, Prize Collecting Arc Routing~\cite{araoz2009clustered}, Prize Collecting Vertex Cover~\cite{karp2009reducibility}, Budgeted Weighted Steiner Tree~\cite{moss2007approximation} \\
\hline
Monotone - Hereditary  & Maximum Planar Subgraph~\cite{junger1994polyhedral}, Maximum Independent Set~\cite{tarjan1977finding},  Maximum Agreement Subtree~\cite{amir1997maximum}, Maximum k-defective clique~\cite{yu2006predicting}, Maximum s-plex problem~\cite{nogueira2020gpu}, Maximum Matching~\cite{galil1986efficient}, Maximum s-bundle graph~\cite{liu2025efficient}, Maximum s-club problem~\cite{schafer2009exact}, Maximum Bipartite subgraph, Maximum k-Vertex Cover~\cite{manurangsi2018note}, Maximum Degree-Bounded Connected Subgraph~\cite{maksimovic2016new}\\
Monotone - Ancestral & Minimum Constraint Removal~\cite{hauser2014minimum}, Minimum Steiner Tree~\cite{kou1981fast}, Minimum Connected Dominating Set~\cite{dai2004extended}, Capacitated Vertex Cover~\cite{guha2003capacitated}, Minimum Feedback Arc Set~\cite{baharev2021exact}, Minimum Equivalent Digraph~\cite{moyles1969algorithm}\\
\hline
\end{tabular}
\vspace{-0.1in}
\end{center}
\vspace{-0.1in}
\end{table}

\vspace{-0.05in}
\subsection{Monotone Problems}

The largest class of graph problems under DoM subgraph optimization problems is the weighted hereditary graph problems. A constraint function is said to be hereditary if all subgraphs of the graph satisfy the constraint given that the graph itself satisfies the constraint. Given a set of positive weighted graph elements, the weighted hereditary problem is to find the subgraph with the maximum total weight that satisfies the given constraint. The weight function of hereditary problems can be made into a DoM objective function by defining the reward function as the sum of the weights of the candidates and the penalty function as $0$ if it satisfies the hereditary constraint and $\infty$ otherwise. 

For example, Hereditary problems such as Maximum Planar Subgraph problems can be solved by $\Delta$Search with the reward function equal to the number of edges and the penalty function $0$ when the subgraph is planar and $\infty$ when the graph is not planar. 
\begin{align*}
    Score(SG) &= R(SG) - P(SG) 
    = |E(SG)| - \begin{cases} 
      0 & \text{if S is planar} \\
       \infty & \text{else} 
   \end{cases}
\end{align*}
Clearly, the above function is a DoM function. As graph elements are added, both the reward and penalty functions increase monotonically. The reward function increases with the number of edges and the penalty function increases monotonically since subgraphs of planar graphs are planar and supergraphs of non-planar graphs are non-planar. The penalty function can be constructed using a linear time Left-Right planarity test~\cite{de2012tremaux}.

\newcommand{\graphwidth}{6cm} 
\newcolumntype{P}[1]{>{\centering\arraybackslash}p{#1}}
\begin{figure}[]
    \centering
    \begin{tabular}{|P{1.1cm}|P{\graphwidth}|P{\graphwidth}|}
        \hline
        \textbf{Graph} & \textbf{Hereditary} & \textbf{Ancestral} \\
        \textbf{element} & & \\
        \hline
        Edge & \includegraphics[width=\graphwidth]{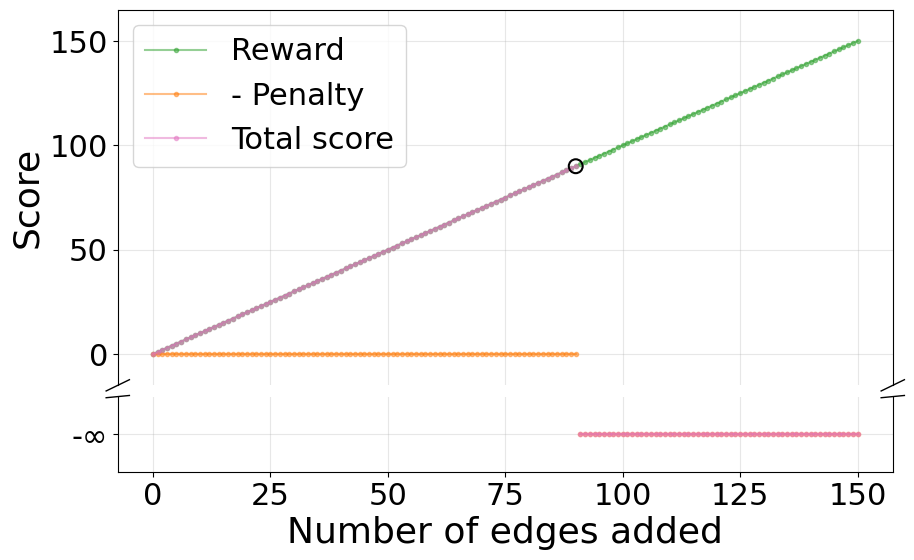} Maximum Planar Subgraph Problem & \includegraphics[width=\graphwidth]{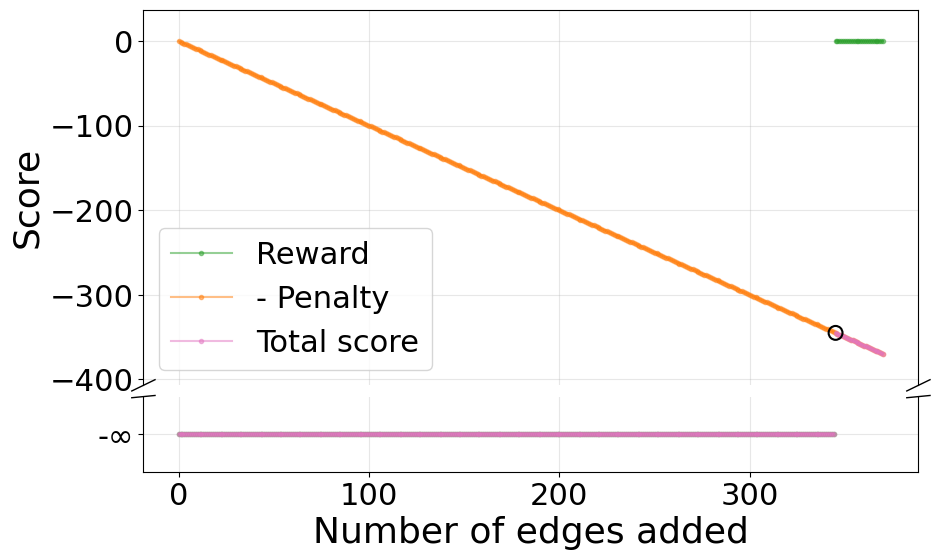} Minimum (weighted) Steiner Tree Problem \\
        \hline
        Vertex & \includegraphics[width=\graphwidth]{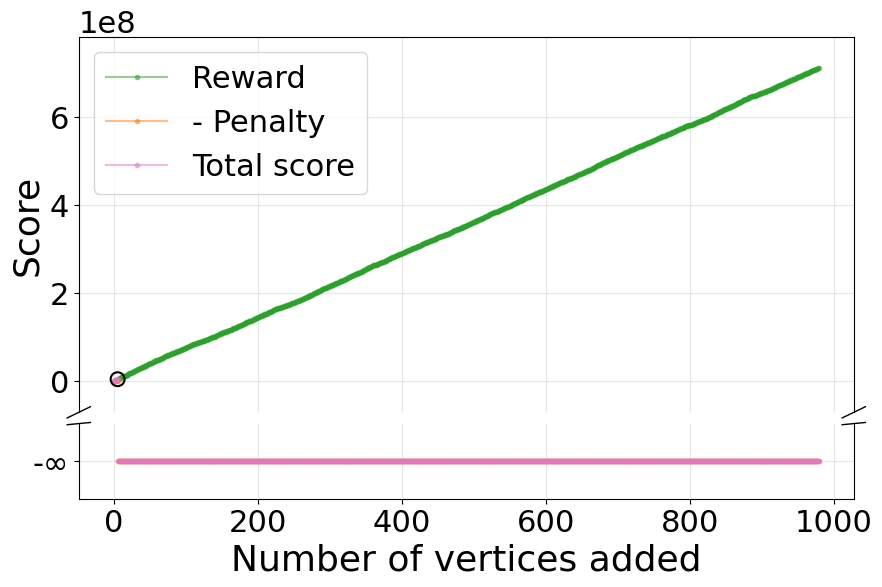} Maximum Weighted Independent Set & \includegraphics[width=\graphwidth]{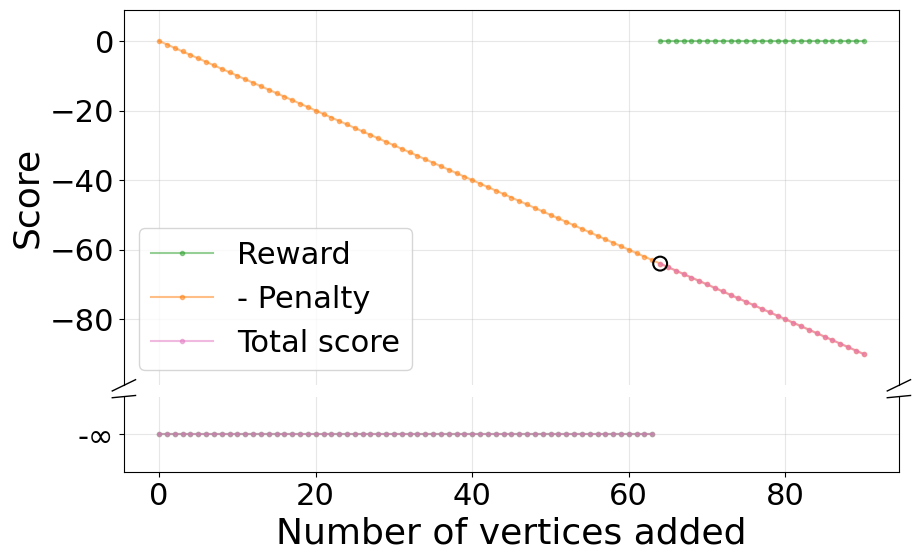} Minimum Connected Dominating Set \\
        \hline
    \end{tabular}
    \caption{The behavior of rewards, negative penalties and total scores of various monotone graph problems with insertion of graph elements in a random ordering.}
    \label{fig:monotone_graph}
    \vspace{-0.15in}
\end{figure}

Consider another hereditary problem: Maximum Weighted Independent Set, where the objective is to find the maximum weight set of vertices that are not adjacent to each other. Unlike the Maximum Planar Subgraph problem, this is an induced subgraph problem (vertex selection) and it is also a weighted problem rather than just selection. This problem can be modeled using DoM as follows:

\vspace{-0.175in}
\begin{align*}
    Score(SG) &= R(SG) - P(SG) 
    =  \Sigma_{v \in V_G(SG)} w(v) - \begin{cases} 
      0 & \text{if }|E_G(SG)| = 0 \\
       \infty & \text{else} 
   \end{cases}
\end{align*}

Ancestral problems which are defined similarly to Hereditary problems can also be formulated easily. A constraint function is said to be ancestral if all super-graphs of a graph satisfy the constraint given that the graph itself satisfies the constraint. For example, the score function of the Minimum (weighted) Steiner tree problem over vertex set $A \subseteq V(G)$ can be defined as follows:
\begin{align*}
    Score(SG) &= R(SG) - P(SG) 
    = \begin{cases} 
      0 & \text{if SG spans over all terminals} \\
       -\infty & \text{else} 
   \end{cases}  - \Sigma_{e \in |E(G)|} w(e)
\end{align*}

In the above Minimum Steiner Tree formulation, the penalty is the total weight of edges since the objective is to find a minimum tree and the reward is $-\infty$ when the subgraph doesn't span over all terminals. Note that there is no constraint to make sure that the resulting graph is a tree because any minimal solution will always be a tree.

Like hereditary problems, induced ancestral subgraph problems can also be formulated in DoM formulation. For example, the Minimum Connected Dominating Set problem can be formulated as:

\vspace{-0.2in}
\begin{align*}
    Score(SG) &= R(SG) - P(SG) 
    = \begin{cases} 
      0 & \text{if $V_G(SG)$ is a connected dominating set} \\
        -\infty & \text{else} 
   \end{cases}  - |V_G(SG)|
\end{align*}

The Hereditary and Ancestral graph problems are together known as the Monotone graph problems. Thus, DoM formulation allows the user to formulate any monotone, weighted or unweighted, subgraph or induced-subgraph problems easily with reward and penalty functions.

As we have seen, the DoM framework is based on the idea that adding graph elements may improve one criterion to be covered while inducing another penalty to be paid. This Reward-Penalty standoff is the basis of the DoM framework. Note that any linear combination of monotone objective functions can be translated into a DoM function by grouping the monotonically increasing functions as the reward function and grouping the monotonically decreasing functions as the penalty function.

This can be seen by considering the following linear combination of monotonically increasing $I_i$s and monotonically decreasing $D_i$s:

\vspace{-0.15in}
\begin{equation}
\begin{split}
    obj(SG) &= \Sigma \alpha_iI_i(SG) + \Sigma \beta_i D_i(SG) + \Sigma_i \gamma_i \\
    &= \underbrace{(\Sigma_{\alpha_i \ge 0} \alpha_iI_i + \Sigma_{\beta_i < 0} \beta_iD_i)}_{\text{Monotonically Increasing Reward}}(SG) - (\underbrace{-\Sigma_{\alpha_i < 0} \alpha_i I_i - \Sigma_{\beta_i \ge 0} \beta_iD_i)}_{\text{Monotonically Increasing Penalty}}(SG) + \underbrace{\Sigma_i \gamma_i}_{\text{Constants that can be ignored}} \\
    &= R(SG) - P(SG)
\end{split}
\label{eq:split}
\end{equation}

\vspace{-0.1in}
The above proof hinges on the two facts that (i) Monotone functions are closed under addition and positive scalar multiplication and (ii) The negation of a monotonically decreasing function is a monotonically increasing function and vice versa. Thus, our formulation can work even in the presence of multiple objective functions, given that all of them are monotone. Given multiple objective functions, the above proof also provides a way to formulate it as a DoM objective function.

\subsection{Non-monotone Problems}
Apart from monotone problems such as Hereditary and Ancestral problems, other intricately scored problems such as Uncapacitated Facility Location Problem (UFLP), Prize Collecting Vertex Cover (PCVC), etc., as can be seen from Table~\ref{table:problems}, can also be modeled by our formulation. In Prize Collecting Vertex Cover Problem, one must find a set of vertices that covers all edges except for the edges that one is willing to be penalized for. Thus, in PCVC, the objective is to find the vertex set with minimum sum of vertex weights and penalized edges.

\vspace{-0.1in}
\begin{align*}
Score(SG) &= - \Sigma_{e \in E_G(G \setminus SG)} w_E(e) - \Sigma_{v \in V_G(SG)} w_V(v)\\
    &= (- \Sigma_{e \in E_G(G \setminus SG)} w_E(e)) - (\Sigma_{v \in V_G(SG)} w_V(v)) \\
&= R(SG) - P(SG)
\end{align*}
\vspace{-0.1in}

As explained, PCVC deals with minimizing the sum of weights of the vertices chosen and the sum of weights of the edges not covered by those vertices. We first convert this to a maximization problem by introducing a negative sign. Then, we apply the formula above to obtain the reward and the penalty functions.

\renewcommand{\graphwidth}{6.8cm} 
\begin{figure}[!t]
    \centering
    \begin{tabular}{|P{\graphwidth}|P{\graphwidth}|}
        \hline
         Uncapacitated Facility Location  & Prize-Collecting Vertex Cover \\
        \hline

        \includegraphics[width=\graphwidth]{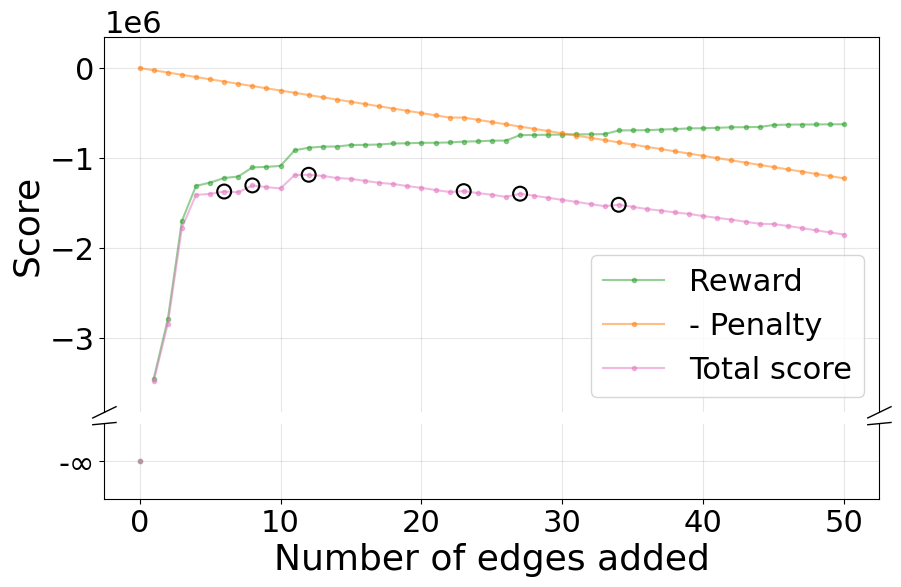} & \includegraphics[width=\graphwidth]{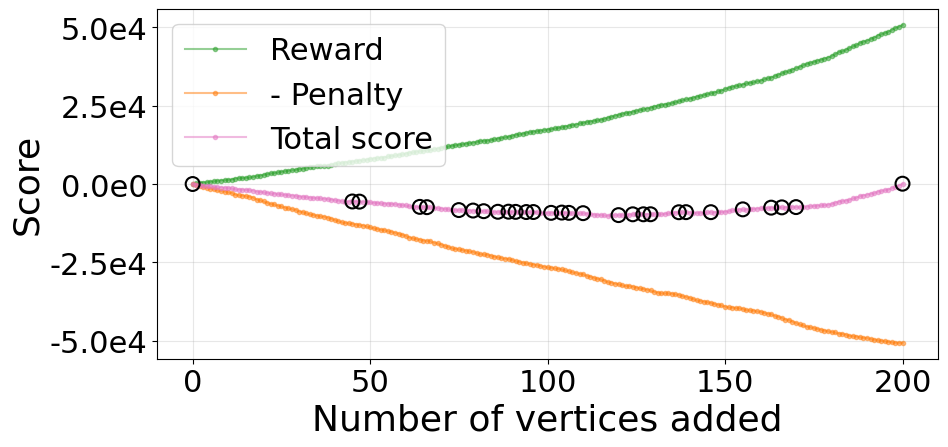}\\
        \hline
    \end{tabular}
    \caption{The behavior of rewards, negation of penalties and total scores of non-monotone DoM graph problems with insertion of vertices in a random ordering.}
    \label{fig:dom_graph}
    \vspace{-0.2in}
\end{figure}

In certain cases, the optimization objective cannot be computed accurately in polynomial time or is not a DoM. In such cases, the objective can be approximated by a DoM function. The performance of an algorithm that relies on this approach will depend on the nature of the approximation. NP-hard problems whose solutions can be verified or scored easily are the best candidates for our approach. 

\begin{table}[!b]
\vspace{-0.15in}
\caption{DoM functions for several Subgraph Extraction Problems and Greedy Action to improve solutions. }
\label{table:scores}
\centering
\small
\begin{tabular}{|c||l||l||c|}
\hline
\textbf{Graph Problem} & $\hspace{0.75in}$\textbf{Reward(SG)} & $\hspace{0.35in}$\textbf{Penalty(SG)} & \textbf{Action} \\
\hline \hline
\makecell{Maximum Planar \\Subgraph} & =$|E_G(SG)|$ &
$ = \begin{cases} 
      0 & \text{if $SG$ is planar}\! \\
      \infty & \text{otherwise} 
   \end{cases} $
& \makecell{Edge \\ Addition}\\ \hline \hline

\makecell{Minimum Steiner \\Tree Problem} & 
$= \begin{cases} 
      -\infty & \text{\makecell{if terminals are not\\ connected in $SG$}} \\
      0 & \text{otherwise} 
   \end{cases} $
& $=\Sigma_{e \in |E_G(SG)|} w(e)$ &\makecell{Edge \\ Deletion}\\ \hline \hline

\makecell{Maximum \\Weighted \\ \!Independent Set} & $=\Sigma_{v \in V_G(SG)}$ w(v) &
\!$ = \begin{cases} 
      0 & \text{if $|E(SG)|=0$}\! \\
      \infty & \text{otherwise} 
   \end{cases} $
&\! \makecell{Vertex \\ Addition}\\ \hline \hline
\makecell{Minimum \\ Connected \\ Dominating Set} & 
$ = \begin{cases} 
      -\infty & \text{\makecell{if $V_G(SG)$ is not a \\connected dominating set}} \\
      0 & \text{otherwise} 
   \end{cases} $
& $=|V_G(SG)|$ &\! \makecell{Vertex \\ Deletion}\\ \hline \hline

\!\multirow{2}{*}{\makecell{Uncapacitated \\ \!\!Facility Location\! \\ Problem}}\!\!& 
 \rule{0pt}{2.5ex} \makecell{$= -\min_{x_{ij}}\Sigma_j \Sigma_i c_{ij}x_{ij}$} $\rule[-1.5ex]{0pt}{0pt}$
 & \makecell{$=\Sigma_i f_i y_i$}&\!\!\ \multirow{2}{*}{\makecell{Vertex \\ \!Addition \&\\  Deletion\\}}\!\!\\
 \cline{2-3}
 & \multicolumn{2}{l||}{\makecell{where $y_i=$ \;\;$1$ if $y_i \in V(SG)$; \;\; $0$ otherwise;   \\$x_{ij} \in \{0, 1\}$ \;\; st \;\; $x_{ij} \le y_i \;\;\&\;\;\; \Sigma_j \; x_{ij}=1$}} &\\
\hline \hline
\makecell{\!Prize Collecting \\Vertex Cover} & 
$ = -\Sigma_{e \in E_G(G \setminus SG)} w_E(e)  $
&\!\!\!\!\makecell{$=\Sigma_{v \in V_G(SG)} w_V(v)$} &\makecell{Vertex  \\ \!Addition \& \\ Deletion}\\ \hline 
\end{tabular}
\end{table}

\section{$\Delta$Search Algorithm}
\label{sec:deltasearch}

Next, we present $\Delta$Search, a heuristic algorithm that can quickly find good solutions for problems that satisfy the DoM formulation. The algorithm only runs the reward and penalty functions at most $O(n^2)$ times, which is much less than many other heuristics. The number of calls to reward and penalty functions reduces to $O(n)$ for monotone graph problems. If needed, this algorithm can be run multiple times with different random seeds to obtain better and diverse solutions. The diversity of the solutions is due to the fixed exploration method of $\Delta$Search which allows it to find different solutions for different orderings of the candidates. This diversity is useful when combined with the exact algorithm to generate multiple diverse solutions for pruning. Next we describe the two key characteristics of $\Delta$Search that allow it to be efficient and general.

\paragraph*{Efficiency.} To address the efficiency of subgraph extraction problems in DoM formulation, we draw inspiration from another high-complexity extraction problem that has achieved remarkable success, namely \emph{delta debugging}~\cite{ddmin,ddmin2}. Given a large failure-inducing input, delta debugging extracts a smaller failure-inducing input to facilitate and simplify debugging by the programmer. Although the number of smaller inputs is exponential, delta debugging employs a binary search-like systematic search strategy that examines a \emph{linear}~\cite{gharachorlu2018avoiding} or \emph{quadratic}~\cite{ddmin, ddmin2} number of smaller inputs and, in practice, delivers impressive results; however, in general, these are suboptimal results. Moreover, it is a black-box technique (i.e., it is not program-specific) that simply runs the program on smaller inputs to find one that is both small and reproduces the failure encountered during the execution on the large input. Thus, delta debugging has similarities to solving non-weighted hereditary problems that are a small subset of DoM problems. We extend both the formulation and the algorithm to bring the benefits of a black-box framework to the wide range of DoM problems.

\paragraph*{Generality.} Due to the monotone nature of the hereditary graph problems, greedy solutions~\cite{chimani2016note, blelloch2012greedy} were heavily explored by prior works. However, a binary search-like approach would achieve a much faster runtime. This observation is the cornerstone for delta debugging, and is what inspired $\Delta$Search. $\Delta$Search also achieves better solution quality than greedy algorithms on average due to its dynamic granularity reduction rather than one-by-one addition of candidates performed by the greedy algorithms. Greedy heuristics for \emph{hereditary} problems can be easily developed by repeatedly adding maximum reward graph elements (i.e., vertices or edges) to an \emph{empty} graph. Similarly, greedy heuristics for \emph{ancestral} problems can be developed by repeatedly deleting maximum penalty graph elements (i.e., vertices or edges) from the \emph{original} graph. These actions for various problems are shown in the last column of Table~\ref{table:scores}. The idea of greedily adding or deleting graph elements for hereditary and ancestral problems respectively motivates our heuristic for DoM to do a search prioritizing the reward for additions and penalties for deletions to achieve a good solution. We call such a heuristic \emph{bidirectional} as it uses both additions and deletions. As a bidirectional search algorithm, it not only generalizes solving hereditary and ancestral graph problems but it is also able to solve other non-monotone DoM problems.

\begin{algorithm}[h]
\caption{\textbf{$\Delta$Search}.}
\label{algo:deltasearch}
\algblockdefx[class]{Class}{EndClass}
[1]{\textbf{class #1}}
[0]{}
\newcommand{\Continue}{\State \textbf{continue} }
\newenvironment{coloredblock}[1][red]{%
\noindent\tcolorbox[colback=#1!10, colframe=#1!10,parbox=false,boxsep=0pt,left=0pt,right=0mm,top=0pt, nobeforeafter]
}{%
\endtcolorbox
}
\algnewcommand{\IIf}[1]{\State\algorithmicif\ #1\ \algorithmicthen}
\algnewcommand{\ElseIIf}[1]{\algorithmicelse\ #1} 
\algnewcommand{\EndIIf}{\unskip\ \algorithmicend\ \algorithmicif}
\algnewcommand{\LineComment}[1]{\State \(\triangleright\) #1}
\small
\begin{algorithmic}[1]

    \Procedure{$\Delta$Search }{ graph, candidates } \label{alg:line:candidates}
        \State $pq \gets MaxPriorityQueue()$ \Comment{Priority-based Worklist}
        \State $next\_pq \gets MaxPriorityQueue()$ \Comment{Holds subsets that cannot improve cur\_graph}
        \State $cur\_graph \gets graph$ \Comment{Current solution}
        \State $pq.insert(~(candidates, DEL, priority=0)~)$ \label{alg:line:del}
        \Statex
        \While{$\neg pq.empty()$}
            \State $(subset, action, priority) \gets pq.extract\_max()$

            \Statex
            \begin{coloredblock}[blue]
            \LineComment{\textcolor{blue}{Prune subset check if possible}} \label{alg:line:prune_start}
            \If{\Call{PruneCondition~}{action, cur\_graph, graph}}
                \State $next\_pq.insert(~(subset,action,priority)~)$
                \Continue
            \EndIf \label{alg:line:prune_end}
            \end{coloredblock}

            \Statex
            \begin{coloredblock}[orange]
            \LineComment{\textcolor{Bittersweet}{Create test\_graph to check score}}
            \State $test\_graph \gets graph.apply(~(subset, action)~)$ \Comment{Add or Delete graph elements} \label{alg:line:apply}
            \State $subset\_priority \gets \Call{Priority~}{test\_graph, action, cur\_graph}$ \Comment{For splitting} \label{alg:line:pri_no}
            \end{coloredblock}

            \Statex
            \begin{coloredblock}[green]
            \LineComment{\textcolor{ForestGreen}{Change cur\_graph if necessary}}
            \State $score\_test\_graph = reward(test\_graph)-penalty(test\_graph)$
            \State $score\_cur\_graph = reward(cur\_graph)-penalty(cur\_graph)$
            \If{$score\_test\_graph > score\_cur\_graph$} \label{alg:line:best_start}
                \State $subset\_priority \gets \Call{Priority~}{cur\_graph, \neg action, test\_graph}$ \Comment{Priority to undo} \label{alg:line:pri_yes}
                \State $cur\_graph \gets test\_graph$ \label{alg:line:best_end}
                \State $pq = pq \cup next\_pq$  \Comment{Current solution changed; validate subsets from next\_pq}
                \State $next\_pq \gets \emptyset$
            \EndIf
            \end{coloredblock}

            \Statex
            \begin{coloredblock}[pink]
            \LineComment{\textcolor{RubineRed}{Split subset into two subsets if possible}}
            \If{$len(subset) == 1$} \Comment{Thus subset cannot improve cur\_graph}
                \State $next\_pq.insert(~(subset, action, priority=subset\_priority)~)$ \label{alg:line:one}
            \Else
                \State $pq.insert(~(subset[:\left\lceil len(subset)/2 \right\rceil], action, priority=subset\_priority)~)$ 
                \State $pq.insert(~(subset[\left\lceil len(subset)/2 \right\rceil:], action, priority=subset\_priority)~)$
            \EndIf
            \end{coloredblock}
            
        \EndWhile

        \Statex
        \State \Return $cur\_graph$
    \EndProcedure
\algstore{myalgorithm}
\end{algorithmic}
\end{algorithm}

\paragraph{Algorithm Details.}
The algorithm for $\Delta$Search is shown in Algorithm~\ref{algo:deltasearch}. Before the search begins, the user provides the graph elements (edges, vertices, or something else) to be operated on as candidates (line. \ref{alg:line:candidates}). The search begins with the entire graph or the entire set of graph elements as the current solution and deletion of all the graph elements from the current solution as the only modification in priority queue. The priority queue always contains modifications (addition or deletion of graph elements) for the current solution ordered by priority. During any iteration, the highest priority modification is chosen to be applied to the current solution (line~\ref{alg:line:apply}). If the solution improves, the modified solution becomes the current solution (lines~\ref{alg:line:best_start}-~\ref{alg:line:best_end}), and the undo of the modification is split and added back to the priority queue to allow for finer search. If the solution doesn't improve, the modification is split and added back to the priority queue for finer search. In either case, if the modification is of only one graph element, then the modification is not split but added directly to the next priority queue (line~\ref{alg:line:one}), since we know that both the modification and its undo have already been tested and therefore cannot improve the current solution.

The priority of the modifications is defined as the potential incremental benefit brought by the action without its cost. That is, for addition, the priority is the potential increase in reward, whereas for deletion, the priority is the potential decrease in penalty as can be seen from Algorithm~\ref{algo:prune}. In the algorithm, the potential incremental benefit of a modification is the reward or penalty change of its parent modification and not the modification itself. Since computing scores is computationally expensive, we reuse the score computation of the parent modification to set priorities for the split modifications. Because applying a modification may or may not improve current solution, the search will either split the undo of the modification or the modification itself into two. Thus, priority is computed for both of these cases on Line~\ref{alg:line:pri_yes} and Line~\ref{alg:line:pri_no} respectively in Algorithm~\ref{algo:deltasearch}. Note that in Line~\ref{alg:line:pri_yes}, the positions of $test\_graph$ and $cur\_graph$ are swapped. This is because the modification converted $cur\_graph$ to $test\_graph$ and therefore, the undo will convert $test\_graph$ (the next $cur\_graph$) to $cur\_graph$.

\begin{algorithm}[!t]
\caption{\textbf{Pruning and Priority}.}
\label{algo:prune}
\small
\begin{algorithmic}[1]
    \algrestore{myalgorithm}
    \Procedure{Priority }{ test\_graph, action, cur\_graph }
        \If{$action == DEL$ }
            \;\;\Return  penalty(cur\_graph) - penalty(test\_graph)
        \ElsIf{$action == ADD$}
            \;\;\Return reward(test\_graph) - reward(cur\_graph)
        \EndIf

    \EndProcedure

    \Statex
    \Procedure{PruneCondition}{action, cur\_graph, graph}
        \If{$action == DEL$}
            \If{$penalty(empty\_graph) == penalty(cur\_graph)$}
                \Return True
            \EndIf
        \ElsIf{$action == ADD$}
            \If{$reward(graph) == reward(cur\_graph)$}
                \Return True
            \EndIf
        \EndIf
        \State \Return False
    \EndProcedure
\end{algorithmic}
\end{algorithm}

Algorithm~\ref{algo:deltasearch} also uses a pruning strategy on Lines~\ref{alg:line:prune_start}-\ref{alg:line:prune_end}. As can be seen from the implementation in Algorithm~\ref{algo:prune}, if the current solution has the same reward as the entire graph, then no addition of graph elements can improve the reward of the current solution and we also know that penalty only increases with graph element additions. Thus, any addition of graph elements can be pruned away when the current solution and the entire graph have the same reward. Similarly, any deletion of graph elements can be pruned away when the penalty of current solution is equal to the penalty of the empty graph. Although this pruning strategy is usually of little benefit to non-monotone problems like PCVC and UFLP, monotone problems can exploit this pruning strategy and non-weighted monotone problems especially can achieve a better complexity of $O(n)$ calls to scoring functions instead of the $O(n^2)$ calls in the general case.

The priority and pruning define the action space used by $\Delta$Search. As we observed in the previous section and in Table~\ref{table:scores}, both addition and deletion may not always be needed. It is enough to use only addition for Maximum Planar Subgraph and Maximum Weighted Independent set. Similarly, it is enough to use only deletion for Ancestral Problems. Thus, in monotone problems, pruning removes unnecessary actions. But problems that are not monotone typically require both additions and deletions. 

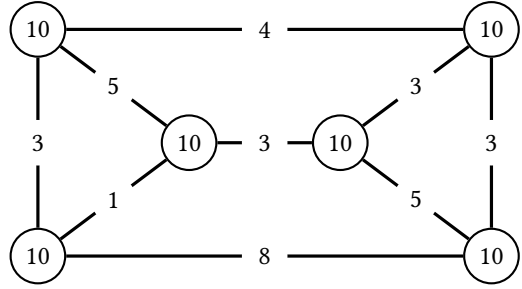
\begin{figure}[!t]
\centering
\vspace{-0.1in}
\begin{tikzpicture}
\begin{scope}[every node/.style={circle,thick,draw}]
    \node (A) at (0,0) {10};
    \node (B) at (2,-1.5) {10};
    \node (C) at (4,-1.5) {10};
    \node (D) at (6, 0) {10};
    \node (E) at (0,-3) {10};
    \node (F) at (6,-3) {10} ;
\end{scope}

\begin{scope}[>={Stealth[black]},
              every node/.style={fill=white,circle},
              every edge/.style={draw=black,very thick}]
    \path [-] (A) edge node {$5$} (B);
    \path [-] (B) edge node {$3$} (C);
    \path [-] (A) edge node {$4$} (D);
    \path [-] (D) edge node {$3$} (C);
    \path [-] (A) edge node {$3$} (E);
    \path [-] (D) edge node {$3$} (F);
    \path [-] (C) edge node {$5$} (F);
    \path [-] (E) edge node {$8$} (F); 
    \path [-] (B) edge node {$1$} (E);
\end{scope}
\end{tikzpicture}
\vspace{0.05in}
\caption{An example of Prize Collecting Vertex Cover with cost of all vertices equal to $10$ and edge penalties shown as edge weights.}
\label{fig:stree}
\vspace{-0.2in}
\end{figure}

\paragraph*{An Example}
An illustration of how $\Delta$Search works is shown in Table~\ref{tab:dsearch}. The example problem is an instance of the PCVC problem with the graph and edge penalties shown in Figure~\ref{fig:stree}. For ease of understanding, let the vertex penalty for all vertices be $10$. Note that $\Delta$Search can work with arbitrary edge and vertex penalties and not just a constant vertex penalty. 

The search in Table~\ref{tab:dsearch} begins with the entire graph as usual. The entire graph has score $-25$ and the first modification, deletion of all graph elements from the current solution, achieves a score $0$ which is better than current solution. Thus, the current solution is set as the empty graph and the addition of all graph elements, which is the undo of deletion of all graph elements, is split into two: Addition of first half of graph elements $\{A, B, C\}$ and Addition of second half of the graph elements $\{C, D, E\}$. These two modifications are then added back into the Priority Queue. The addition of graph elements $\{A, B, C\}$ does not achieve a score better than empty graph, hence the modification itself is split into two: Addition of graph element $\{A\}$ and Addition of graph elements $\{B, C\}$. 

The addition of graph element $\{A\}$ achieves a better solution and hence current solution (empty graph) with $\{A\}$ is set as the current solution. Note that the undo of addition of $\{A\}$ to the current graph is not added back to the priority queue since we have already tested the empty graph and know that it is inferior to the current solution. Thus, the undo of this modification, the Deletion of $\{A\}$ is only added to the $next\_pq$. The $next\_pq$ containing Deletion of $\{A\}$ merges with the priority queue only after the current solution changes in iteration $8$. 

Table~\ref{tab:dsearch} only shows the search up to $9$ iterations. After this, the search proceeds with testing each of the modification in the priority queue which only produces inferior solutions, adding nothing back to the priority queue. The priority queue thus becomes empty and the search terminates with $\{A, D\}$ as the solution.

\paragraph*{Complexity}
Let us now analyze the time complexity of $\Delta$Search in terms of the number of calls to the scoring functions. Let us first look at the number of modifications that will be considered during the run of $\Delta$Search.  Since an $n$ element list can only be split into two $n-1$ times, the number of modifications considered by the algorithm is at most $O(n)$, where $n$ is the number of graph elements. Therefore, at any instant during the $\Delta$Search, the priority queue will only have at most $O(n)$ elements. The algorithm terminates if there is no element in priority queue. Thus, the current solution must change every $O(n)$ iterations to refill the priority queue or it terminates. Since every current solution change invalidates at least one modification, the number of times current solution changes is at most $O(n)$. Since there are at most $O(n)$ current solution changes and since there are at most $O(n)$ calls to scoring functions between each current solution change, the total number of calls to the scoring functions during the runtime of the algorithm is at most $O(n^2)$.


This analysis is less tight for monotone problems as these problems work with only one action due to pruning. Here, all modifications can only be applied once or forever be discarded. Thus, hereditary and ancestral problems call the scoring functions at most $O(n)$ times.

\begin{longtable}[!t]{|c|P{1.2cm}|P{1.1cm}|P{0.8cm}|c|P{1.0cm}|P{0.6cm}|P{2.7cm}|}
\caption{Illustration of Space Exploration by $\Delta$Search for the Prize-Collecting Vertex Cover Problem with weight of every vertex equal to $10$. The elements in the priority queue are of the form (subset, action, priority).}
\label{tab:dsearch}
\\ \hline
&{\small\textbf{Solution}}  & {\small\textbf{Test}} &{\small\textbf{Action}}&  {\small\textbf{Test Graph}} & {\small\textbf{R(SG) }} & {\small\textbf{Score}} & {\small\textbf{Worklist}}\\
&{\small{\textit{cur\_graph}}}  & \textit{subset} &  & \textit{test\_graph} &  {\small\textbf{-P(SG)}}& & \textit{pq}\\
\hline \hline
- & \{A, B, C, D, E, F\} & -  & -  &

\scalebox{0.55}{
\parbox[c]{14em}{

\begin{tikzpicture}
\begin{scope}[every node/.style={circle,thick,draw,fill=GreenYellow}]
    \node (A) at (0,0) {A};
    \node (B) at (1.5,-1.5) {E};
    \node (C) at (3,-1.5) {F};

    \node (D) at (4.5, 0) {B};
    \node (E) at (0,-3) {C};
    \node (F) at (4.5,-3) {D} ;
\end{scope}

\begin{scope}[>={Stealth[black]},
              every node/.style={fill=white,circle},
              every edge/.style={draw=ForestGreen,very thick}]
    \path [-] (A) edge node {$5$} (B);
    \path [-] (B) edge node {$3$} (C);
    \path [-] (A) edge node {$4$} (D);
    \path [-] (D) edge node {$3$} (C);
    \path [-] (A) edge node {$3$} (E);
    \path [-] (D) edge node {$3$} (F);
    \path [-] (C) edge node {$5$} (F);
    \path [-] (E) edge node {$8$} (F); 
    \path [-] (B) edge node {$1$} (E);
\end{scope}
\end{tikzpicture}
}
}
& 35 - 60 & -25 & \{(\{A, B, C, D, E, F\}, DEL, 0) \} \\
1 & \{A, B, C, D, E, F\}  & \{A, B, C, D, E, F\}  & DEL &

\scalebox{0.55}{
\parbox[c]{14em}{
\begin{tikzpicture}
\begin{scope}[every node/.style={circle,thick,draw}]
    \node (A) at (0,0) {A};
    \node (B) at (1.5,-1.5) {E};
    \node (C) at (3,-1.5) {F};
    \node (D) at (4.5, 0) {B};
    \node (E) at (0,-3) {C};
    \node (F) at (4.5,-3) {D} ;
\end{scope}

\begin{scope}[>={Stealth[black]},
              every node/.style={fill=white,circle},
              every edge/.style={draw=black,very thick}]
    \path [-] (A) edge node {$5$} (B);
    \path [-] (B) edge node {$3$} (C);
    \path [-] (A) edge node {$4$} (D);
    \path [-] (D) edge node {$3$} (C);
    \path [-] (A) edge node {$3$} (E);
    \path [-] (D) edge node {$3$} (F);
    \path [-] (C) edge node {$5$} (F);
    \path [-] (E) edge node {$8$} (F); 
    \path [-] (B) edge node {$1$} (E);
\end{scope}
\end{tikzpicture}
}
}
& 0 - 0 & 0 & \{(\{A,B,C\},ADD,-35), (\{D, E, F\},ADD,-35)\}\\
2 & \{\}& \{A,B,C\}& ADD &

\scalebox{0.55}{
\parbox[c]{14em}{
\begin{tikzpicture}
\begin{scope}[every node/.style={circle,thick,draw}]
    \node (F) at (4.5,-3) {D} ;
    \node (B) at (1.5,-1.5) {E};
    \node (C) at (3,-1.5) {F};
\end{scope}

\begin{scope}[every node/.style={circle,thick,draw,fill=GreenYellow}]
    \node (A) at (0,0) {A};
    \node (D) at (4.5, 0) {B};
    \node (E) at (0,-3) {C};
\end{scope}

\begin{scope}[>={Stealth[black]},
              every node/.style={fill=white,circle},
              every edge/.style={draw=ForestGreen,very thick}]
    \path [-] (E) edge node {$8$} (F); 
    \path [-] (A) edge node {$5$} (B);
    \path [-] (A) edge node {$4$} (D);
    \path [-] (D) edge node {$3$} (C);
    \path [-] (A) edge node {$3$} (E);
    \path [-] (D) edge node {$3$} (F);
    \path [-] (B) edge node {$1$} (E);
\end{scope}

\begin{scope}[>={Stealth[black]},
              every node/.style={fill=white,circle},
              every edge/.style={draw=black,very thick}]
    \path [-] (B) edge node {$3$} (C);
    \path [-] (C) edge node {$5$} (F);
\end{scope}
\end{tikzpicture}
}
}
& 27 - 30 & -3 &\{ (\{A\}, ADD, 27), (\{B, C\}, ADD, 27), (\{D, E, F\}, ADD, 0)\}\\
3 & \{\}&\{A\}& ADD &

\scalebox{0.55}{
\parbox[c]{14em}{
\begin{tikzpicture}
\begin{scope}[every node/.style={circle,thick,draw}]
    \node (B) at (1.5,-1.5) {E};
    \node (C) at (3,-1.5) {F};
    \node (D) at (4.5, 0) {B};
    \node (E) at (0,-3) {C};
    \node (F) at (4.5,-3) {D} ;
\end{scope}

\begin{scope}[every node/.style={circle,thick,draw,fill=GreenYellow}]
    \node (A) at (0,0) {A};
\end{scope}

\begin{scope}[>={Stealth[black]},
              every node/.style={fill=white,circle},
              every edge/.style={draw=ForestGreen,very thick}]
    \path [-] (A) edge node {$5$} (B);
    \path [-] (A) edge node {$4$} (D);
    \path [-] (A) edge node {$3$} (E);
\end{scope}

\begin{scope}[>={Stealth[black]},
              every node/.style={fill=white,circle},
              every edge/.style={draw=black,very thick}]
    \path [-] (B) edge node {$3$} (C);
    \path [-] (D) edge node {$3$} (C);
    \path [-] (D) edge node {$3$} (F);
    \path [-] (C) edge node {$5$} (F);
    \path [-] (E) edge node {$8$} (F); 
    \path [-] (B) edge node {$1$} (E);
\end{scope}
\end{tikzpicture}
}
}
& 12 - 10 & 2 &\{ (\{B, C\}, ADD, 27), (\{D, E, F\}, ADD, 0)\}\\ 
4 & \{A\} & \{B, C\}& ADD& 

\scalebox{0.55}{
\parbox[c]{14em}{
\begin{tikzpicture}
\begin{scope}[every node/.style={circle,thick,draw}]
    \node (F) at (4.5,-3) {D} ;
    \node (B) at (1.5,-1.5) {E};
    \node (C) at (3,-1.5) {F};
\end{scope}

\begin{scope}[every node/.style={circle,thick,draw,fill=GreenYellow}]
    \node (A) at (0,0) {A};
    \node (D) at (4.5, 0) {B};
    \node (E) at (0,-3) {C};
\end{scope}

\begin{scope}[>={Stealth[black]},
              every node/.style={fill=white,circle},
              every edge/.style={draw=ForestGreen,very thick}]
    \path [-] (E) edge node {$8$} (F); 
    \path [-] (A) edge node {$5$} (B);
    \path [-] (A) edge node {$4$} (D);
    \path [-] (D) edge node {$3$} (C);
    \path [-] (A) edge node {$3$} (E);
    \path [-] (D) edge node {$3$} (F);
    \path [-] (B) edge node {$1$} (E);
\end{scope}

\begin{scope}[>={Stealth[black]},
              every node/.style={fill=white,circle},
              every edge/.style={draw=black,very thick}]
    \path [-] (B) edge node {$3$} (C);
    \path [-] (C) edge node {$5$} (F);
\end{scope}
\end{tikzpicture}
}
}
& 27 - 30 & -3 & \{ (\{B\}, ADD, 15), (\{C\}, ADD, 15), (\{D,~E, F\}, ADD, 0)\}\\
5 & \{A\} & \{B\}& ADD &

\scalebox{0.55}{
\parbox[c]{14em}{
\begin{tikzpicture}
\begin{scope}[every node/.style={circle,thick,draw}]
    \node (E) at (0,-3) {C};
    \node (F) at (4.5,-3) {D} ;
    \node (B) at (1.5,-1.5) {E};
    \node (C) at (3,-1.5) {F};
\end{scope}

\begin{scope}[every node/.style={circle,thick,draw,fill=GreenYellow}]
    \node (A) at (0,0) {A};
    \node (D) at (4.5, 0) {B};
\end{scope}

\begin{scope}[>={Stealth[black]},
              every node/.style={fill=white,circle},
              every edge/.style={draw=ForestGreen,very thick}]
    \path [-] (A) edge node {$5$} (B);
    \path [-] (A) edge node {$4$} (D);
    \path [-] (D) edge node {$3$} (C);
    \path [-] (A) edge node {$3$} (E);
    \path [-] (D) edge node {$3$} (F);
\end{scope}

\begin{scope}[>={Stealth[black]},
              every node/.style={fill=white,circle},
              every edge/.style={draw=black,very thick}]
    \path [-] (B) edge node {$3$} (C);
    \path [-] (C) edge node {$5$} (F);
    \path [-] (E) edge node {$8$} (F); 
    \path [-] (B) edge node {$1$} (E);
\end{scope}
\end{tikzpicture}
}
}
& 18 - 20 & -2 & \{ (\{C\}, ADD, 15), (\{D, E, F\}, ADD, 0)\}\\ 

6 & \{A\}& \{C\}& ADD & 

\scalebox{0.55}{
\parbox[c]{14em}{
\begin{tikzpicture}
\begin{scope}[every node/.style={circle,thick,draw}]
    \node (F) at (4.5,-3) {D} ;
    \node (B) at (1.5,-1.5) {E};
    \node (C) at (3,-1.5) {F};
    \node (D) at (4.5, 0) {B};
\end{scope}

\begin{scope}[every node/.style={circle,thick,draw,fill=GreenYellow}]
    \node (A) at (0,0) {A};
    \node (E) at (0,-3) {C};
\end{scope}

\begin{scope}[>={Stealth[black]},
              every node/.style={fill=white,circle},
              every edge/.style={draw=ForestGreen,very thick}]
    \path [-] (E) edge node {$8$} (F); 
    \path [-] (A) edge node {$5$} (B);
    \path [-] (A) edge node {$4$} (D);
    \path [-] (A) edge node {$3$} (E);
    \path [-] (B) edge node {$1$} (E);
\end{scope}

\begin{scope}[>={Stealth[black]},
              every node/.style={fill=white,circle},
              every edge/.style={draw=black,very thick}]
    \path [-] (B) edge node {$3$} (C);
    \path [-] (C) edge node {$5$} (F);
    \path [-] (D) edge node {$3$} (C);
    \path [-] (D) edge node {$3$} (F);
\end{scope}
\end{tikzpicture}
}
}
& 21 - 20 & 1 & \{ (\{D,E,F\}, ADD, 0) \}\\ 

7 & \{A\}& \{D, E, F\} & ADD &

\scalebox{0.55}{
\parbox[c]{14em}{
\begin{tikzpicture}
\begin{scope}[every node/.style={circle,thick,draw}]
    \node (D) at (4.5, 0) {B};
    \node (E) at (0,-3) {C};
\end{scope}

\begin{scope}[every node/.style={circle,thick,draw,fill=GreenYellow}]
    \node (A) at (0,0) {A};
    \node (F) at (4.5,-3) {D} ;
    \node (B) at (1.5,-1.5) {E};
    \node (C) at (3,-1.5) {F};
\end{scope}

\begin{scope}[>={Stealth[black]},
              every node/.style={fill=white,circle},
              every edge/.style={draw=ForestGreen,very thick}]
    \path [-] (A) edge node {$5$} (B);
    \path [-] (A) edge node {$4$} (D);
    \path [-] (A) edge node {$3$} (E);
    \path [-] (B) edge node {$3$} (C);
    \path [-] (D) edge node {$3$} (C);
    \path [-] (D) edge node {$3$} (F);
    \path [-] (C) edge node {$5$} (F);
    \path [-] (E) edge node {$8$} (F); 
    \path [-] (B) edge node {$1$} (E);
\end{scope}

\begin{scope}[>={Stealth[black]},
              every node/.style={fill=white,circle},
              every edge/.style={draw=black,very thick}]
\end{scope}
\end{tikzpicture}
}
}
& 35 - 40 & -5 & \{ (\{D\}, ADD, 23), (\{E, F\}, ADD, 23) \}\\ 

\rowcolor{LimeGreen!20} 8 & \{A\} & \{D\} & ADD &

\scalebox{0.55}{
\parbox[c]{14em}{
\begin{tikzpicture}
\begin{scope}[every node/.style={circle,thick,draw}]
    \node (D) at (4.5, 0) {B};
    \node (E) at (0,-3) {C};
    \node (B) at (1.5,-1.5) {E};
    \node (C) at (3,-1.5) {F};
\end{scope}

\begin{scope}[every node/.style={circle,thick,draw,fill=GreenYellow}]
    \node (A) at (0,0) {A};
    \node (F) at (4.5,-3) {D} ;
\end{scope}

\begin{scope}[>={Stealth[black]},
              every node/.style={fill=white,circle},
              every edge/.style={draw=ForestGreen,very thick}]
    \path [-] (A) edge node {$5$} (B);
    \path [-] (A) edge node {$4$} (D);
    \path [-] (A) edge node {$3$} (E);
    \path [-] (D) edge node {$3$} (F);
    \path [-] (C) edge node {$5$} (F);
    \path [-] (E) edge node {$8$} (F); 
\end{scope}

\begin{scope}[>={Stealth[black]},
              every node/.style={fill=white,circle},
              every edge/.style={draw=black,very thick}]
    \path [-] (B) edge node {$3$} (C);
    \path [-] (D) edge node {$3$} (C);
    \path [-] (B) edge node {$1$} (E);
\end{scope}
\end{tikzpicture}
}
}
& 28 - 20 & 8 & \{ (\{E, F\}, ADD, 23), (\{A\}, DEL, 10), (\{D\},~DEL, 10),  (\{C\},~ADD, 9),  (\{B\},~ADD, 6) \}\\ 
9 & \{A, D\} & \{E, F\} & ADD &

\scalebox{0.55}{
\parbox[c]{14em}{
\begin{tikzpicture}
\begin{scope}[every node/.style={circle,thick,draw}]
    \node (D) at (4.5, 0) {B};
    \node (E) at (0,-3) {C};
    \node (F) at (4.5,-3) {D} ;
\end{scope}

\begin{scope}[every node/.style={circle,thick,draw,fill=GreenYellow}]
    \node (A) at (0,0) {A};
    \node (B) at (1.5,-1.5) {E};
    \node (C) at (3,-1.5) {F};
\end{scope}

\begin{scope}[>={Stealth[black]},
              every node/.style={fill=white,circle},
              every edge/.style={draw=ForestGreen,very thick}]
    \path [-] (A) edge node {$5$} (B);
    \path [-] (A) edge node {$4$} (D);
    \path [-] (A) edge node {$3$} (E);
    \path [-] (B) edge node {$3$} (C);
    \path [-] (D) edge node {$3$} (C);
    \path [-] (C) edge node {$5$} (F);
    \path [-] (B) edge node {$1$} (E);
\end{scope}

\begin{scope}[>={Stealth[black]},
              every node/.style={fill=white,circle},
              every edge/.style={draw=black,very thick}]
    \path [-] (D) edge node {$3$} (F);
    \path [-] (E) edge node {$8$} (F); 
\end{scope}
\end{tikzpicture}
}
}
& 35 - 40 & -5 & \{ (\{A\}, DEL, 10), (\{D\},~DEL, 10),  (\{C\},~ADD,~9), (\{E\},~ADD, 7), (\{F\},~ADD, 7),  (\{B\},~ADD, 6) \}\\ 
\vdots & \vdots & \vdots & \vdots & \vdots & \vdots & \vdots & \vdots \\
\hline
\end{longtable}

\section{Implementation and Programming Interface}
\label{sec:interface}

$\Delta$Search is implemented as a Python library around three input components: (i) Graph - the graph whose subgraph is to be extracted; (ii) Graph element (Vertex or Edge) - which defines the changes allowed to the graph; and (iii) Score functions (Reward and Penalty). The mutual independence among these three components allows their reuse for different problems.

\begin{figure}[!t]
    \centering
    \includegraphics[width=0.8\linewidth]{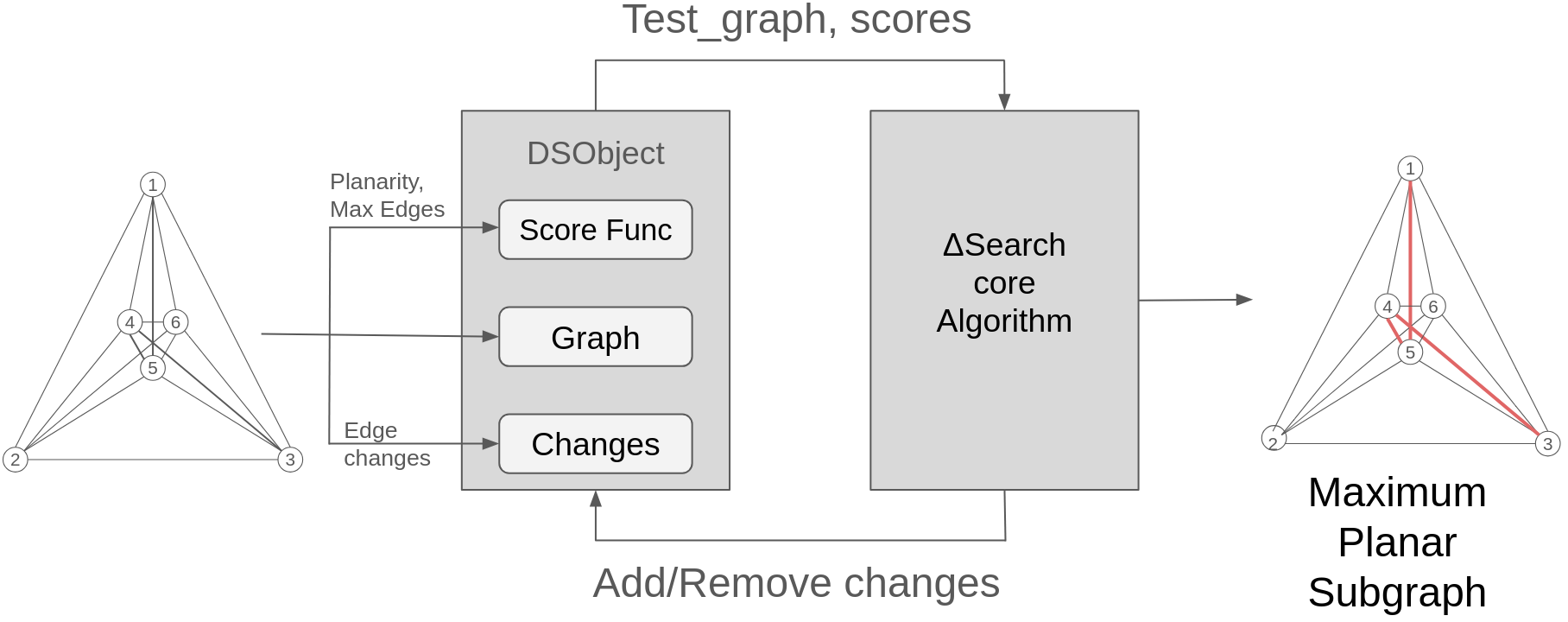}
    \vspace{0.05in}
    \caption{The $\Delta$Search architecture.}
    \label{fig:framework}
    \vspace{-0.2in}
\end{figure}

\vspace{-0.1in}
\paragraph*{Graph Elements}
The graph elements represent the smallest changes that can be made to the graph. Examples of graph elements include vertices for induced subgraph problems and edges for subgraph problems. Since vertices and edges are the most common graph element types, they are implemented in the library. The user is also allowed to create custom graph elements. For example, let us consider the Minimum Constraint Removal (MCR) Problem. In the MCR problem, given a graph, a start vertex, an end vertex and a list of subsets of the vertex set called obstacles, one must find the path between the start and end vertex that minimizes the number of obstacles it intersects. In this problem, the graph element is the obstacles (subsets of vertex sets) and with it, $\Delta$Search will search the smallest set of obstacles to delete that still retains reachability between start and end vertex. To write custom graph elements, the user only has to define the number of graph elements and how to add and remove them.

\vspace{-0.1in}
\paragraph*{Scoring Functions}
The $\Delta$Search framework accepts an array of functions that sum up to the optimization objective (Reward functions and negation of Penalty functions). These functions are assigned as reward or penalty automatically by $\Delta$Search using the logic presented in Equation~\ref{eq:split}. Since $\Delta$Search calls these functions multiple times, it is important to optimize these functions well. In fact, optimizing the scoring functions is the only tuning left to the user. This is in contrast with various other meta-heuristics where the user is supposed to tune the hyper-parameters based on the user's understanding of the algorithm. On the other hand, in $\Delta$Search, the user only has to implement highly optimized scoring functions. In addition to the three inputs, the $\Delta$Search framework consists of two modules. The first one is the DSObject - a wrapper around the inputs and the second one is the core algorithm.

\vspace{-0.1in}
\paragraph*{DSObject}
The DSObject is a wrapper around all three inputs. The DSObject provides methods to add and remove subset modifications and to compute reward and penalty. The core search algorithm of $\Delta$Search calls these methods during its search. Although we have shown the modification and scoring done directly on the graph in Algorithm~\ref{algo:deltasearch}, the actual implementation uses the DSObject abstraction so that any data structure can be used with $\Delta$Search. The DSObject abstractions also allow for better optimizations such as incremental scoring (to avoid computing the scores from scratch every time), caching of scores for reuse and sharing of computation between reward and penalty functions. However, it may also introduce dependence between the scoring functions and graph elements (such as a scoring function constrained to work with only one type of graph element). Thus, users should avoid directly using the DSObject interface unless necessary. More details on the uses and implementation can be found in our code repository.

\vspace{-0.1in}
\paragraph*{The Core Algorithm}
As can be seen from Figure~\ref{fig:framework}, the core algorithm controls DSObject by assigning the modification to apply next and the DSObject applies the modification and computes the score after the modification has been applied. The main reason for the separation of DSObject and the core algorithm is to ensure that the core algorithm remains graph-agnostic so that it could even be applied to other problems.

\section{Experimental Evaluation}
\label{sec:experiments}

We evaluate the effectiveness of $\Delta$Search by solving multiple graph problems and comparing its solution quality and execution time against problem-specific heuristic baselines on standard graph datasets, as shown in Table~\ref{tab:placeholder}. We use an AMD EPYC 7713 processor with a memory limit of $50$ GB. $\Delta$Search is implemented in Python 3.10 with the networkx 2.8.8 library. Heathcliff~\cite{healthcliff} (MST), MMWIS~\cite{mmwis} (MWIS), GuMWIS~\cite{gumwis} (MWIS), BinCSA~\cite{bincsa} (UFLP), and APBEA~\cite{apbea} (UFLP) use their original C++ implementations and were compiled with gcc 12.3.0 with -O3 optimization flag. All remaining baselines are implemented in Python 3.10.

\begin{table}[!h]
    \centering
    \vspace{-0.1in}
    \caption{Baseline algorithms and Datasets for evaluation of different DoM problem instances.}
    \label{tab:placeholder}
    \small
    \begin{tabular}{|c||c||c|}
        \hline
        \textbf{Problem} & \textbf{Baselines} & \textbf{Graph Datasets} \\
        \hline
        \hline
        Maximum Planar Subgraph & Naive, Cactus+~\cite{chimani2009non}& \!North~\cite{di2000drawing}, Steinlib~\cite{koch2001steinlib}\!\\
        \hline
        \!Minimum Steiner Tree Problem\! & TM heuristic~\cite{takahashi1980approximate}, Heathcliff~\cite{bonnet2018pace} & Steinlib~\cite{koch2001steinlib}\\
        \hline
         & GWMIN, GWMAX, GWMIN2~\cite{sakai2003note} & \multirow{3}{*}{VR~\cite{dong2021new}}\\
        Maximum Weighted & MMWIS~\cite{grossmann2023finding}, GUMWIS~\cite{gu2021towards},  & \\
        Independent Set& Local Ratio (LR), Fractional LR~\cite{bar2004local} & \\
        \hline
        Minimum Connected& GR\_CDS, GR\_CDS\_pruned~\cite{fu2016greedy}, & \multirow{2}{*}{LPNMR~\cite{jovanovic2013ant}}\\
         Dominating Set  &  IC\_MIS\_ST~\cite{sun2019minimum} & \\
         \hline
        \!Uncapacitated Facility Location\! & BinCSA~\cite{sonucc2021binary}, APBEA~\cite{sonucc2023adaptive} & ORLIB~\cite{beasley1990or}, M*~\cite{kratica2001solving}\\
        \hline
        Prize Collecting Vertex Cover & Local Ratio~\cite{bar2004local} & Synthetic~\cite{milanovic2010solving}\\
        \hline
    \end{tabular}
\end{table}

\vspace{-1.25em}
\subsection{Case Studies}

\subsubsection{Maximum Planar Subgraph}

Given a graph $G = (V,E)$, the Maximum Planar Subgraph (MPS) problem is to find the largest edge subset $F \subseteq E$ such that $SG$, the graph induced by $F$, is planar. The MPS problem is NP-hard~\cite{cualinescu1998better}, which led to the rise in popularity of heuristic algorithms for MPS. Approximate algorithms for Maximum Planar Subgraph Problem have applications in graph drawing. The Planarization method~\cite{chimani2009non}, one of the strongest heuristics to draw a graph with fewest crossings starts with a planar subgraph and adds other edges incrementally.

Multiple heuristics have been developed for the Maximum Planar Subgraph Problem. \cite{chimani2016note} presents a list of heuristics: Boyer and Myrvold method (BM), Cactus Algorithm (C) and the Naive method (Ni) developed for finding large planar subgraphs and evaluates them against each other. The simplest of the methods, Naive, adds edges one by one to find a maximal planar subgraph. To ensure that these methods find maximal and not just large ones, we can postprocess the other algorithms using the Naive method to achieve maximality (BM+, Cactus+) as done in \cite{chimani2016note}. Since \cite{chimani2016note} shows that BM, BM+ and Cactus are inferior to Cactus+ in solution quality, we remove them from our evaluations.

The cactus algorithm is a 7/18-approximation algorithm whereas our $\Delta$Search and the Naive algorithm are both 1/3-approximation algorithms. However in experiments, both $\Delta$Search and the Naive algorithm outperform the cactus algorithm in almost all graphs. Cactus+, on the other hand, is competitive with $\Delta$Search and the naive algorithm. We use the non-planar graphs of the established real-world datasets from North~\cite{di2000drawing} (423 instances) and SteinLib~\cite{koch2001steinlib} (586 instances). 

\begin{table}[]
    \vspace{-0.1in}
    \centering
    \caption{MPS - $\Delta$Search vs. Naive and Cactus+ algorithms~\cite{chimani2009non}. ↑ is better for quality and time.}
    \label{tab:planar_imp}
    \small
    \begin{tabular}{|c||c|c||c|c|}
        \hline
        \multirow{2}{*}{\textbf{Algorithm}} & \multicolumn{2}{c||}{\textbf{Improvement for North}~\cite{di2000drawing}} & \multicolumn{2}{c|}{\textbf{Improvement for Steinlib}~\cite{koch2001steinlib}} \\
        \cline{2-5}
        & \textbf{Avg Quality} & \textbf{Avg Time} & \textbf{Avg Quality} & \textbf{Avg Time} \\
        \hline
        Naive & +0.54\% & -22.03\% & +0.53\% & -20.56\% \\
        Cactus+ & -0.18\% & -2.17\% & -0.35\% & -1.27\% \\
        \hline
    \end{tabular}
\end{table}
\begin{figure}[!t]
    \centering
    \includegraphics[width=0.8\linewidth]{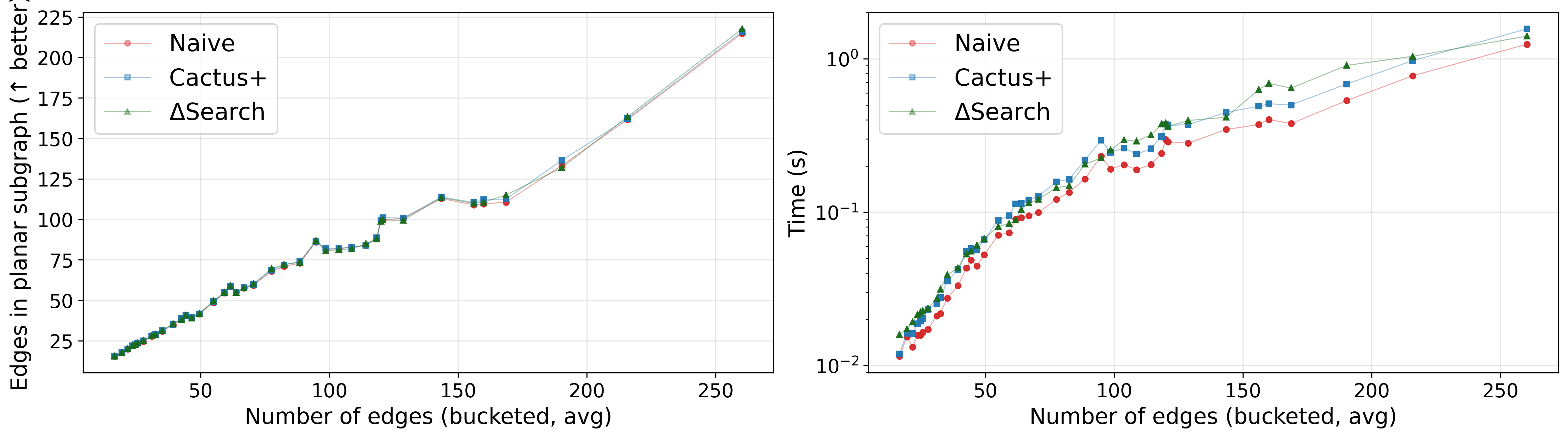}
    \caption{Maximum Planar Subgraph - $\Delta$Search vs. Cactus and Cactus+ baselines for the North~\cite{di2000drawing} dataset.}
    \label{fig:planar_quality}
    \vspace{-0.1in}
\end{figure}

Figure~\ref{fig:planar_quality} shows the solution and time performance of Naive, Cactus+ and $\Delta$Search for the graph instances, and Table~\ref{tab:planar_imp} shows the average improvement in solution quality and average runtime speedup of $\Delta$Search with respect to Naive and Cactus+ for the North and Steinlib datasets. Since the combined number of graph instances is $1009$, as shown in Figure~\ref{fig:planar_quality}, we bucket the graph instances into $40$ buckets based on the number of edges. We can see that on average, $\Delta$Search produces better solutions than the Naive greedy algorithm with $0.54\%$ improvement. $\Delta$Search is also competitive against the Cactus+ algorithm in solution quality and runtime performance. Thus, $\Delta$Search is competitive against the algorithms developed for the Maximum Planar Subgraph despite being a general framework.

\subsubsection{Minimum Steiner Tree Problem}

The Minimum (Weighted) Steiner Tree problem is another important NP-hard problem in combinatorial optimization. It plays a central role in integrated circuit design, network design and facility location~\cite{ljubic2021solving}. Given a graph $G$ and a subset of vertices called terminals $A \subseteq V(G)$, the Minimum Weighted Steiner Tree problem is to find the subgraph with minimum weighted set of edges such that all the terminals are reachable to each other in the subgraph.

Multiple works have been developed to solve the Steiner Tree problem. \cite{ljubic2021solving} presents a survey on the recent advances in solving Steiner trees. We compare our $\Delta$Search framework with the winner of the PACE 2018 Track C challenge~\cite{bonnet2018pace} on the Steiner Tree problem, which we will refer to as Heathcliff. Note that we use the original implementation of Heathcliff in C++ from \cite{healthcliff}. We also compare our work against a heuristic algorithm named TM-heuristic~\cite{takahashi1980approximate} which achieves an approximation ratio of $2-2/|A|$ of the minimum solution. We use Steinlib~\cite{koch2001steinlib}, an established real-world benchmark for the Steiner tree problem, for our evaluations. Since Heathcliff's runtime is much higher than TM-heuristic and $\Delta$Search, we limit its runtime to $300$ seconds for all the Steiner tree experiments.

\begin{table}[]
    \centering
    \caption{MST - $\Delta$Search vs. TM and Heathcliff algorithms.↑ is better for quality and time.}
    \label{tab:steiner_imp}
    \small
    \begin{tabular}{|c||c|c|}
        \hline
        \multirow{2}{*}{\textbf{Algorithm}} & \multicolumn{2}{c|}{\textbf{Improvement for Steinlib}~\cite{koch2001steinlib}} \\
        \cline{2-3}
        & \textbf{Avg Quality} & \textbf{Avg Time} \\
        \hline
        TM Heuristic & -36.58\% & -90.98\% \\
        Heathcliff & -31.11\% & +170796.17\% \\
        \hline
    \end{tabular}
    \vspace{-0.1in}
\end{table}

\begin{figure}
    \centering
    \includegraphics[width=0.8\linewidth]{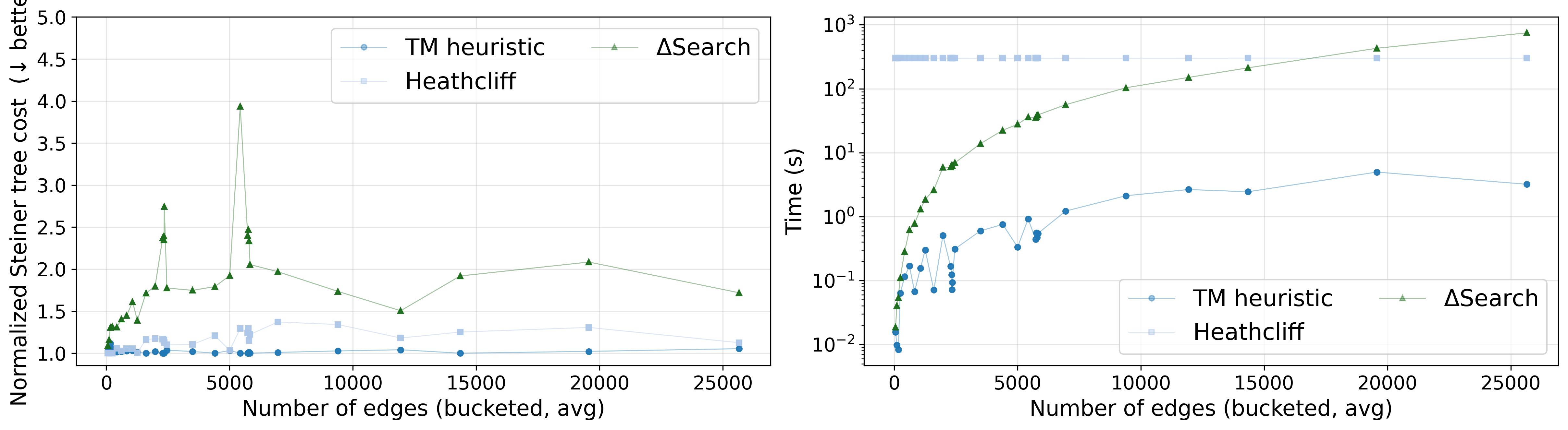}
    \caption{Minimum Steiner Tree - $\Delta$Search vs. TM~\cite{takahashi1980approximate} and Heathcliff~\cite{bonnet2018pace} baselines on Steinlib~\cite{koch2001steinlib} dataset.}
    \label{fig:steiner_quality}
    \vspace{-0.15in}
\end{figure}

Figure~\ref{fig:steiner_quality} shows the cost of the solutions returned by $\Delta$Search, TM Heuristic and Heathcliff normalized by the best solution produced and the runtimes of all three algorithms. Figure~\ref{fig:steiner_quality} and Table~\ref{tab:steiner_imp} show that unlike Maximum Planar Subgraph Problem, $\Delta$Search performs poorly in solution quality compared to both the algorithms. This is not unexpected as both of these algorithms were developed for the Steiner problem. Interestingly, Heathcliff performs worse than TM Heuristic even with its huge runtime. A bigger time limit might be beneficial for Heathcliff. The poorer runtime performance of $\Delta$Search with respect to TM heuristic is mainly due to the fact that TM heuristic is constructive in nature. That is, unlike $\Delta$Search which checks the validity of subsets multiple times, TM heuristic produces a valid Steiner tree by construction. Hence, it is much faster than $\Delta$Search. On the other hand, Heathcliff uses an evolutionary algorithm which is much slower than $\Delta$Search. Though $\Delta$Search is not comparable to TM heuristic, it could still be competitive with Heathcliff if run multiple times to match the runtime. The mechanism of running $\Delta$Search multiple times and the effectiveness of this approach is shown in Section~\ref{sec:repeat}.

\subsubsection{Maximum Weighted Independent Set}

The Maximum Weighted Independent Set (MWIS) problem is yet another NP-hard problem which deals with finding a set of vertices that are not adjacent and whose total weight is maximum. Maximum Weighted Independent Set problem has applications in image segmentation~\cite{brendel2010segmentation} and multi-object tracking~\cite{brendel2011multiobject} in Computer Vision and transmission scheduling in networks.

Multiple greedy~\cite{sakai2003note} (GWMIN, GWMAX and GWMIN2) heuristics and other heuristics ~\cite{gu2021towards} (GUMWIS) and metaheuristic~\cite{grossmann2023finding} (MMWIS) works have been developed for computing Maximum Weighted Independent Set. We use all the above works as baseline for $\Delta$Search for Maximum Weighted Independent Set Problem. Note that the original C++ implementation of gumwis and mmwis from \cite{gumwis} and  \cite{mmwis} are used for comparison. We also compare our work against general frameworks Local Ratio and Fractional Local Ratio as they too can work with this problem.

A significant amount of research has been devoted to reduction techniques for the Weighted Independent Set problem. \cite{grossmann2024comprehensive} presents a survey on multiple reduction techniques developed and used by prior work on Weighted Independent Set problems. Since gumwis uses multiple reduction techniques internally, to keep comparisons fair, we apply reductions to all graphs before running the baseline algorithms. For datasets, we use the dataset VR instances~\cite{dong2021new} from Vehicle Routing application.

\begin{table}[]
    \centering
    \caption{MWIS - $\Delta$Search vs. other MWIS algorithms~\cite{sakai2003note,bar2004local,gumwis,mmwis}. ↑ is better for quality and time.}
    \label{tab:mwis_imp}
    \small
    \begin{tabular}{|c||c|c|}
        \hline
        \multirow{2}{*}{\textbf{Algorithm}} & \multicolumn{2}{c|}{\textbf{Improvement for VR}~\cite{dong2021new}} \\
        \cline{2-3}
        & \textbf{Avg Quality} & \textbf{Avg Time} \\
        \hline
        GWMIN & -10.10\% & -83.97\% \\
        GWMAX & -9.15\%& +204.76\% \\
        GWMIN2 & -9.75\% & +47.26\% \\
        Local Ratio & -5.38\% & -92.6\% \\
        Fractional Local Ratio & -8.15\% & -75.63\% \\
        GUMWIS & -10.12\% & -95.63\% \\
        MMWIS & -12.39\% & +2718.00\% \\
        \hline
    \end{tabular}
    \vspace{-0.15in}
\end{table}

\begin{figure}
    \centering
    \includegraphics[width=0.8\linewidth]{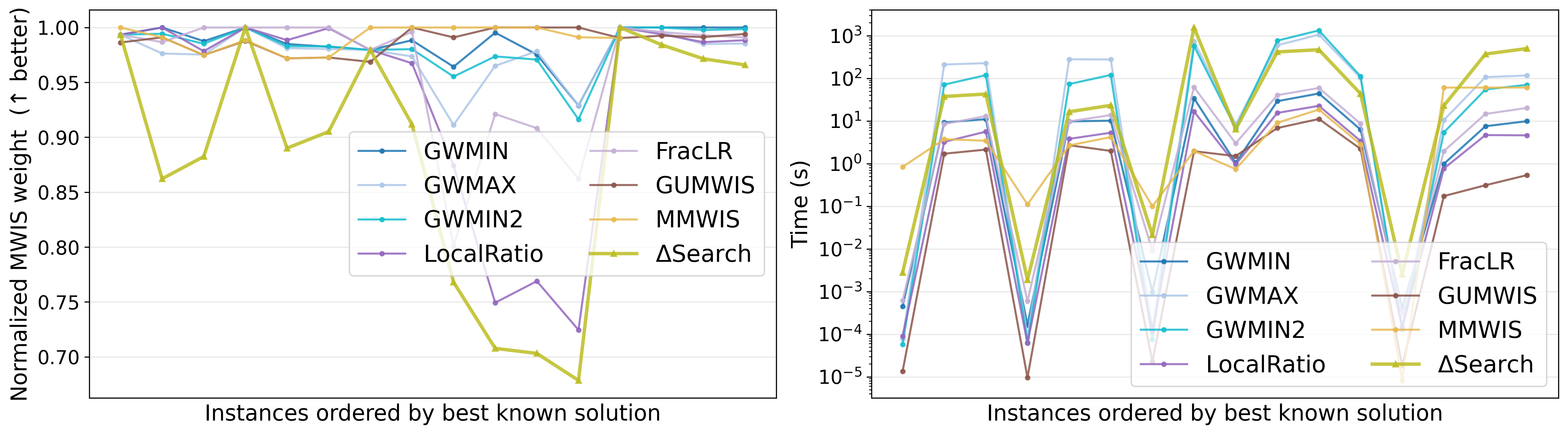}
    \caption{Maximum Weight Independent Set - $\Delta$Search vs. various baselines for VR~\cite{dong2021new} graph instances.}
    \label{fig:mwis_quality}
    \vspace{-0.15in}
\end{figure}

Figure~\ref{fig:mwis_quality} illustrates the sum of edge weight of solutions computed by the MWIS algorithms normalized by the best found solution and the runtime of the MWIS algorithms. Table~\ref{tab:mwis_imp} shows the ratio of solution produced and runtime by $\Delta$Search compared to the other MWIS algorithms. GWMIN, Local Ratio and Fractional Local Ratio are faster than the $\Delta$Search algorithm due to being construction-based algorithms. Although $\Delta$Search is comparable to Local Ratio, due to its user provided algorithm for solving smaller instances, it is able to achieve better solution quality compared to Local Ratio. $\Delta$Search achieves around $90$\% of the solution quality of the custom algorithms despite not having any user-provided guidance.

\subsubsection{Minimum Connected Dominating Set}

The Minimum Connected Dominating Set (MCDS) is an NP-hard problem with applications in constructing backbones in ad hoc and wireless networks~\cite{sun2019minimum, fu2016greedy}. We use the prior works ICIK and ICML from \cite{sun2019minimum} and GR\_CDS from \cite{fu2016greedy} as baselines for this experiment. For dataset, we use the instances from Tenth
International Conference on Logic Programming and Nonmonotonic Reasoning (LPNMR’09)~\cite{jovanovic2013ant}.

\begin{table}[]
    \centering
    \caption{MCDS - $\Delta$Search vs. other MCDS algorithms.↑ is better for quality and time.}
    \label{tab:codo_imp}
    \small
    \begin{tabular}{|c||c|c|}
        \hline
        \multirow{2}{*}{\textbf{Algorithm}} & \multicolumn{2}{c|}{\textbf{Improvement for LPNMR}~\cite{jovanovic2013ant}} \\
        \cline{2-3}
        & \textbf{Avg Quality} & \textbf{Avg Time} \\
        \hline
        GR\_CDS & -21.62\% & -99.13\% \\
        GR\_CDS\_pruned & -22.27\% & -99.13\% \\
        IC\_MIS\_ST & +44.90\% & -61.51\% \\
        \hline
    \end{tabular}
    \vspace{-0.1in}
\end{table}

\begin{figure}
    \centering
    \includegraphics[width=0.8\linewidth]{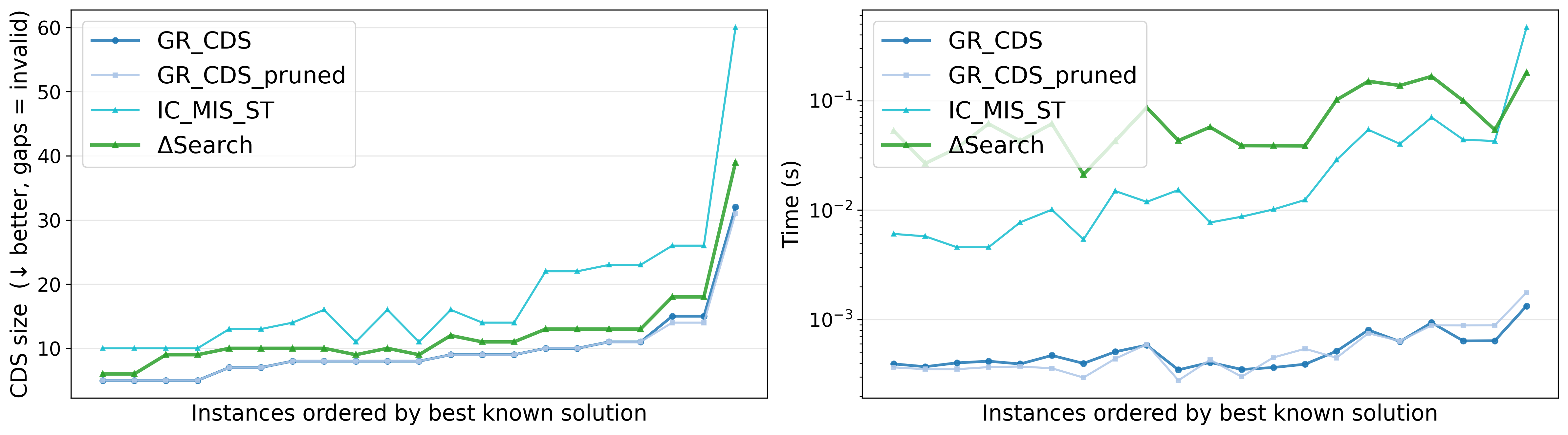}
    \caption{Minimum Connected Dominating Set - $\Delta$Search vs. other MCDS baselines for the LPNMR~\cite{jovanovic2013ant} dataset.}
    \label{fig:connected_dominating_quality}
    \vspace{-0.15in}
\end{figure}

Figure~\ref{fig:connected_dominating_quality} gives the number of edges in the solutions for GR\_CDS, GR\_CDS\_pruned, IC\_MIS\_ST and $\Delta$Search and their runtimes and Table~\ref{tab:codo_imp} shows the improvement of $\Delta$Search over the MCDS baselines. From the figure, we observe that $\Delta$Search achieves better solution than IC\_MIS\_ST, albeit being slower than the latter. But again, $\Delta$Search falls behind in runtime as all the baselines are constructive. However, it is still noteworthy that it achieved around $80$\% of the solution of the best custom algorithms GR\_CDS and GR\_CDS\_pruned and also being able to perform better than one existing custom algorithm IC\_MIS\_ST.

\begin{table}[!b]
    \vspace{-0.15in}
    \centering
    \caption{UFLP - Non-optimal solutions found by the UFLP algorithms for ORLIB and M* datasets. \textcolor{red}{*} indicates crashes of the publicly available baseline implementations of APBEA\cite{apbea} and BinCSA\cite{bincsa} on larger instances.}
    \label{tab:ucfl_nonoptimal}
    \small
    \begin{tabular}{|c|c|c|c|}
        \hline
         \textbf{Instance} & \textbf{BinCSA} & \textbf{APBEA} & \textbf{$\Delta$Search} \\
         \hline
         MO3 & \textcolor{red}{1521.473} & 1516.773 & 1516.773 \\
         MR1 & \textcolor{red}{2609.08} & 2608.148 & 2608.148329 \\
         MR3 & 2788.25 & 2788.25 & \textcolor{red}{2793.324183} \\
         MS1 & \textcolor{red}{$*$} & \textcolor{red}{$*$} & 5283.757394\\
         MT1 & \textcolor{red}{$*$} & \textcolor{red}{$*$} & 10069.802769\\
         \hline
    \end{tabular}
    \vspace{-0.15in}
\end{table}

\begin{figure}[!b]
    \centering
    \includegraphics[width=0.8\linewidth]{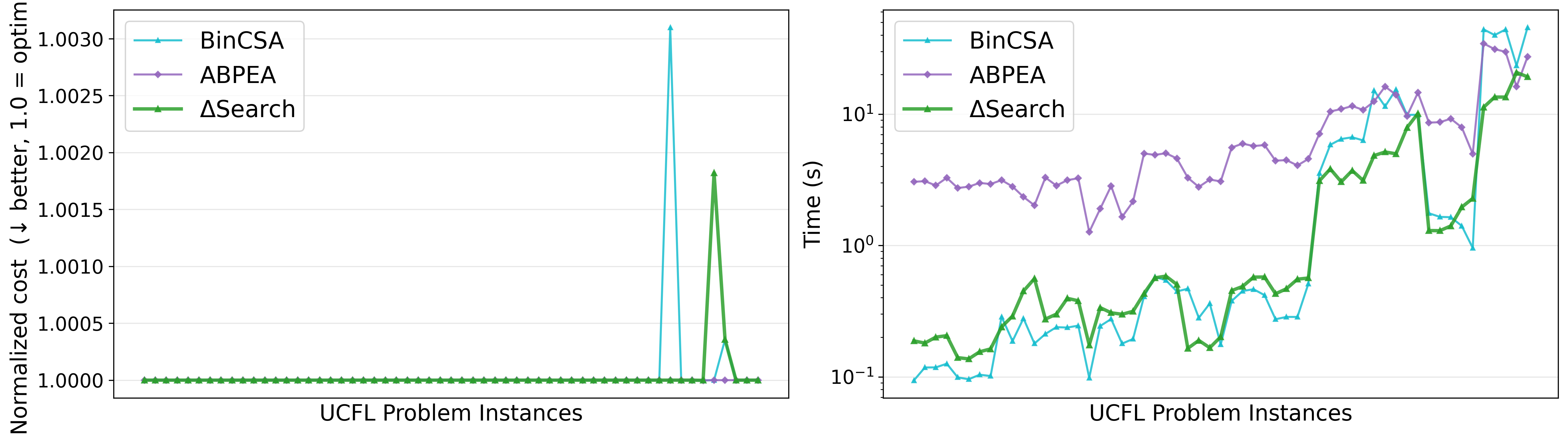}
    \caption{Uncapacitated Facility Location - $\Delta$Search vs. APBEA~\cite{sonucc2023adaptive} and BinCSA~\cite{sonucc2021binary} baselines for ORLIB~\cite{beasley1990or} and M*~\cite{kratica2001solving} datasets. The solution costs are normalized using the optimal solution cost.}
    \label{fig:ucfl_quality}
\end{figure}

\subsubsection{Uncapacitated Facility Location Problem}

UFLP is one of the most famous NP-hard problems~\cite{cornuejols1983uncapicitated} with a wide range of applications from resource allocation, network architecture and computer vision to infrastructure construction of schools, hospitals, warehouses etc~\cite{zhang2023fast}.

An UFLP problem deals with finding a set of locations to setup factories to minimize total cost. A UFLP instance consists of a set of  consumer locations and a set of locations where factories can be set up. There exists two costs in an UFLP instance: (i) Transportation cost for delivering services to the customers from any of the factory locations and (ii) Set up cost for setting up factories in a location. There is no limit on how many customers a factory can serve but all the customers must be served. The objective of an UFLP instance is to minimize the total cost or sum of transportation cost and the setup cost. Mathematically, given $n$ factory locations and $m$ customers, the UFLP instance can be described as follows:

\vspace{-0.15in}
{\small
\begin{align*}
    \min_{x_{ij}, y_i} \text{  } & \Sigma_j \Sigma_i c_{ij}x_{ij} + \Sigma_i f_i y_i \\
\text{where } &  \Sigma_i x_{ij}=1, &\forall j=1,2, \cdots m \\
    & x_{ij} \le y_i, &\forall i=1,2, \cdots n, \forall j=1,2, \cdots m \\
    & x_{ij}, y_i \in \{0, 1\}, &\forall i=1,2, \cdots n, \forall j=1,2, \cdots m \\
\end{align*}
}
\vspace{-0.35in}

\noindent
where $c_{ij}$ is the transportation cost of service from factory $i$ to customer $j$, $f_i$ is the set up cost for setting up a factory at location $i$. In $\Delta$Search, we model this problem as a selection problem where the locations where the factories need to be setup are selected. The assignment of factories to customers is then done by the reward function by assigning the closest factory to any customer. The scoring functions can be seen in Table~\ref{table:scores}.

Due to its high popularity, various heuristics and meta-heuristics have been constructed for the UFLP problem. In this work, for UFLP, we compared $\Delta$Search against the greedy algorithm from \cite{guha1999greedy} and metaheuristics APBEA, EGTOA and BinCSA from \cite{sonucc2023adaptive,zhang2023fast, sonucc2021binary}. APBEA is considered the state of the art algorithm for UFLP. Figure~\ref{fig:ucfl_quality} shows the performance of $\Delta$Search compared against APBEA and BinCSA. The greedy algorithm and EGTOA algorithm were removed from the figure as they produced multiple non-optimal results. Since the time taken by APBEA and BinCSA were very high compared to $\Delta$Search, to make comparisons fair, we ran $\Delta$Search $100$ times and chose the best result. Details of how running $\Delta$Search multiple times results in better solution is explained in the next subsection. The original C++ implementation of APBEA and BinCSA from \cite{apbea} and \cite{bincsa} respectively are used for comparison. For datasets, we use ORlib~\cite{beasley1990or}, the most well known dataset in this area with 15 instances and M*~\cite{kratica2001solving}, another UFLP dataset with 22 instances. 

Figure~\ref{fig:ucfl_quality} illustrates the cost of solution computed by and runtime of BinCSA, APBEA and $\Delta$Search algorithms. From Figure~\ref{fig:ucfl_quality}, we can see that all the algorithms achieve the optimal solution. Table~\ref{tab:ucfl_nonoptimal} shows the instances where the three algorithms fail to achieve optimal solutions. $\Delta$Search fails to achieve optimal solution only for one instance whereas BinCSA fails to achieve optimal solution for two instances. Moreover, both BinCSA and APBEA suffer from a segfault when ran on MS1 and MT1 as they store the computational data on the stack which causes a stack overflow for larger instances. Thus, $\Delta$Search is superior in performance to both BinCSA and APBEA. Figure~\ref{fig:ucfl_quality} also shows that $\Delta$Search is faster than both APBEA and BinCSA even though APBEA and BinCSA were implemented in C++ while $\Delta$Search is implemented in Python. Thus, $\Delta$Search is on par in quality and runtime with the state of the art algorithm APBEA even with a slower implementation.

\subsubsection{Prize Collecting Vertex Cover}

Prize Collecting Vertex Cover (PCVC) is another example of a non-monotone subgraph problem which is NP-hard. It was first introduced in \cite{karp2009reducibility}. Although it does not have many applications by itself, many of its variants are well researched~\cite{zhou2024approximation, liu2023primal, markarian2021algorithmic}. The main reason we chose PCVC to evaluate our framework is because it is a non-monotone graph problem and that PCVC can be solved using Local Ratio. Hence, Local Ratio will be the only baseline for this experiment. For benchmark, we use the instance generation algorithm~\cite{milanovic2010solving} from the related generalized vertex cover problem.

\begin{table}[]
    \centering
    \caption{PCVC - $\Delta$Search vs. Local Ratio for the synthetic dataset~\cite{milanovic2010solving}.↑ is better for quality and time.}
    \label{tab:pcvc_imp}
    \small
    \begin{tabular}{|c||c|c|}
        \hline
        \multirow{2}{*}{\textbf{Algorithm}} & \multicolumn{2}{c|}{\textbf{Ratio for Synthetic}~\cite{milanovic2010solving} dataset} \\
        \cline{2-3}
        & \textbf{Avg Quality} & \textbf{Avg Time} \\
        \hline
        Local Ratio & +38.19\% & -75.78\% \\
        \hline
    \end{tabular}
    \vspace{-0.15in}
\end{table}

\begin{figure}
    \centering
    \includegraphics[width=0.8\linewidth]{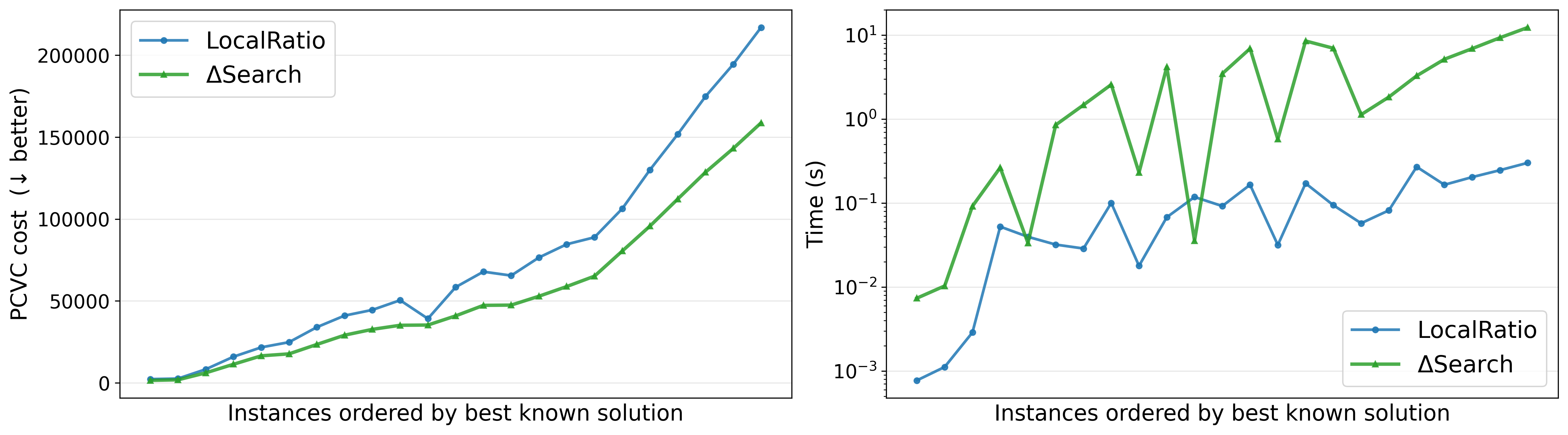}
    \caption{Prize Collecting Vertex Cover - $\Delta$Search vs. Local Ratio baselines for synthetic dataset~\cite{milanovic2010solving}.}
    \label{fig:pcvc_quality}
    \vspace{-0.15in}
\end{figure}

Figure~\ref{fig:pcvc_quality} illustrates the cost of solutions and runtimes of Local Ratio and $\Delta$Search algorithms. Table~\ref{tab:pcvc_imp} shows the total average ratio of the PCVC cost and runtimes between $\Delta$Search and Local Ratio. As can be seen from Figure~\ref{fig:pcvc_quality} and Table~\ref{tab:pcvc_imp}, $\Delta$Search achieves better solution quality compared to Local Ratio. On the other hand, Local Ratio achieves a better runtime than $\Delta$Search. Local Ratio is a constructive algorithm which only generates solutions that improve the overall solution. On the other hand, $\Delta$Search runs the scoring function multiple times which slows down the $\Delta$Search algorithm. Yet despite guidance from the user, Local Ratio fails to find a better solution than $\Delta$Search.

\vspace{-0.5em}

\subsection{Multi-Start $\Delta$Search}
\label{sec:repeat}

\begin{figure}[!h]
    \vspace{-0.1in}
    \centering
    \includegraphics[width=0.8\linewidth]{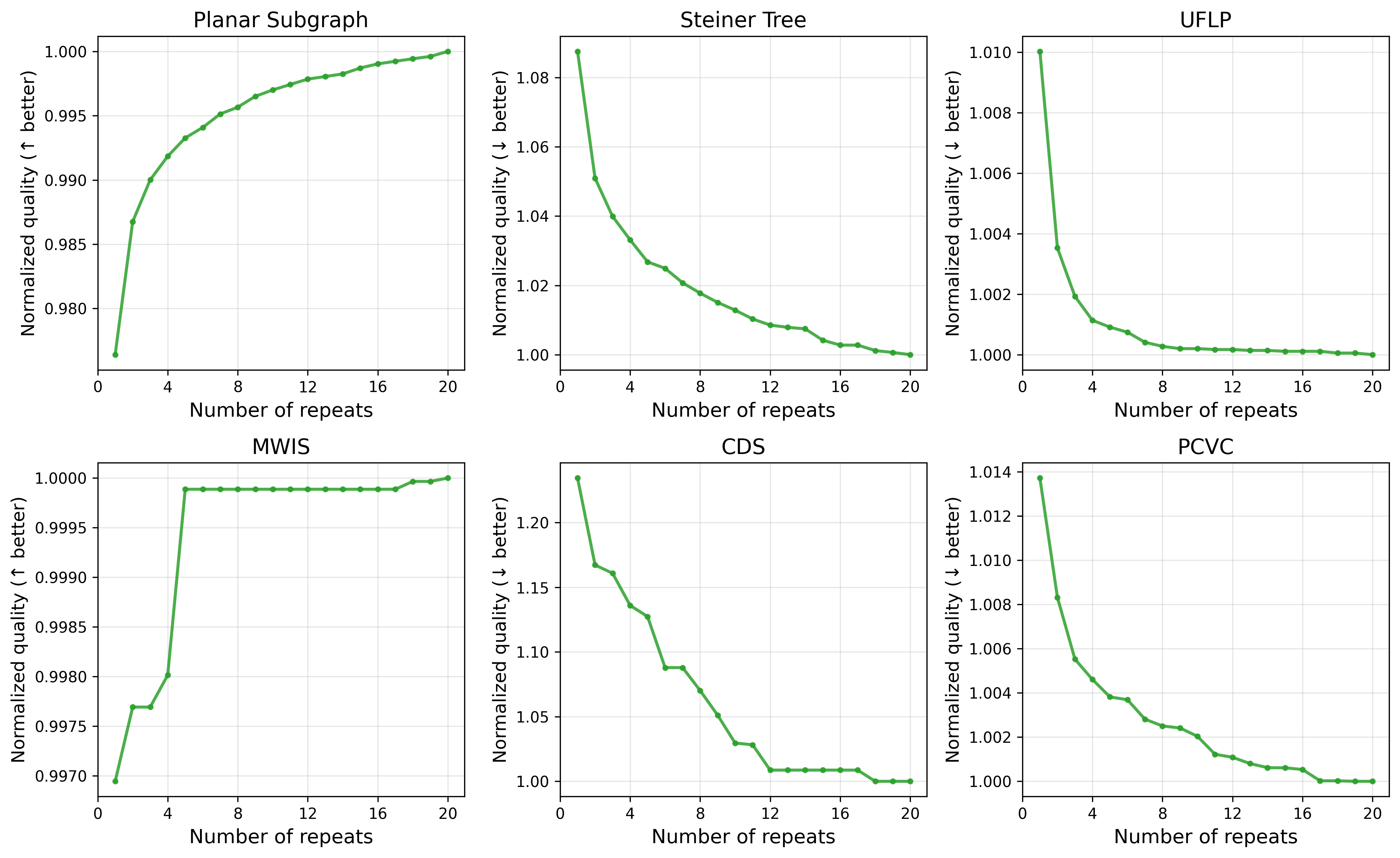}
    \caption{Improvement in solution quality for different problems with increasing number of repeats.}
    \label{fig:repeat}
    \vspace{-0.1in}
\end{figure}

$\Delta$Search is a deterministic algorithm that returns the same solution when provided the same initial candidates for the graph elements. But $\Delta$Search is indeed sensitive to the ordering of the graph elements. This sensitivity can be exploited by running the algorithm multiple times with different orderings. Each repeat of the algorithm will change the search space of the algorithm, improving the chances of achieving a better solution at least in one of the repeats. This is a common technique in combinatorial optimization problems known as Multi-start. Figure~\ref{fig:repeat} shows the improvement in solution quality with an increasing number of repeats.

Figure~\ref{fig:repeat} shows that the quality of the solution improves with the number of repeats. The improvement in solution quality usually decreases with the number of repeats and eventually converges to a steady state. There exists a sweet spot that balances the solution quality and the runtime that depends on the nature of the problem and the size of the instance.

\vspace{-0.5em}
\subsection{Accelerating Exact Search using $\Delta$Search}
\label{sec:deltaexact}

In \cite{trukhanov2013algorithms}, an exact exponential algorithm, Russian Doll Search (RDS), was introduced to solve weighted hereditary problems. The main motivation was the pruning mechanisms that hereditary problems allowed for when searching via RDS. This modified RDS uses bounds generated by its previous searches to prune its future searches. We integrate $\Delta$Search into this algorithm, as shown in Figure~\ref{fig:overview}, to improve the bounds so that the modified RDS prunes more search space thereby achieving a good speedup. Inspired by n-way parallelism~\cite{cledat2011efficiently}, we run $3$ threads of $\Delta$Search in parallel to generate bounds for the modified RDS. The $\Delta$Search threads run based on the exploration done by the exact algorithm. That is, the $\Delta$Search threads will only search for solutions that the exact algorithm has not explored. 

\begin{figure}
    \centering
    \includegraphics[width=0.5\linewidth]{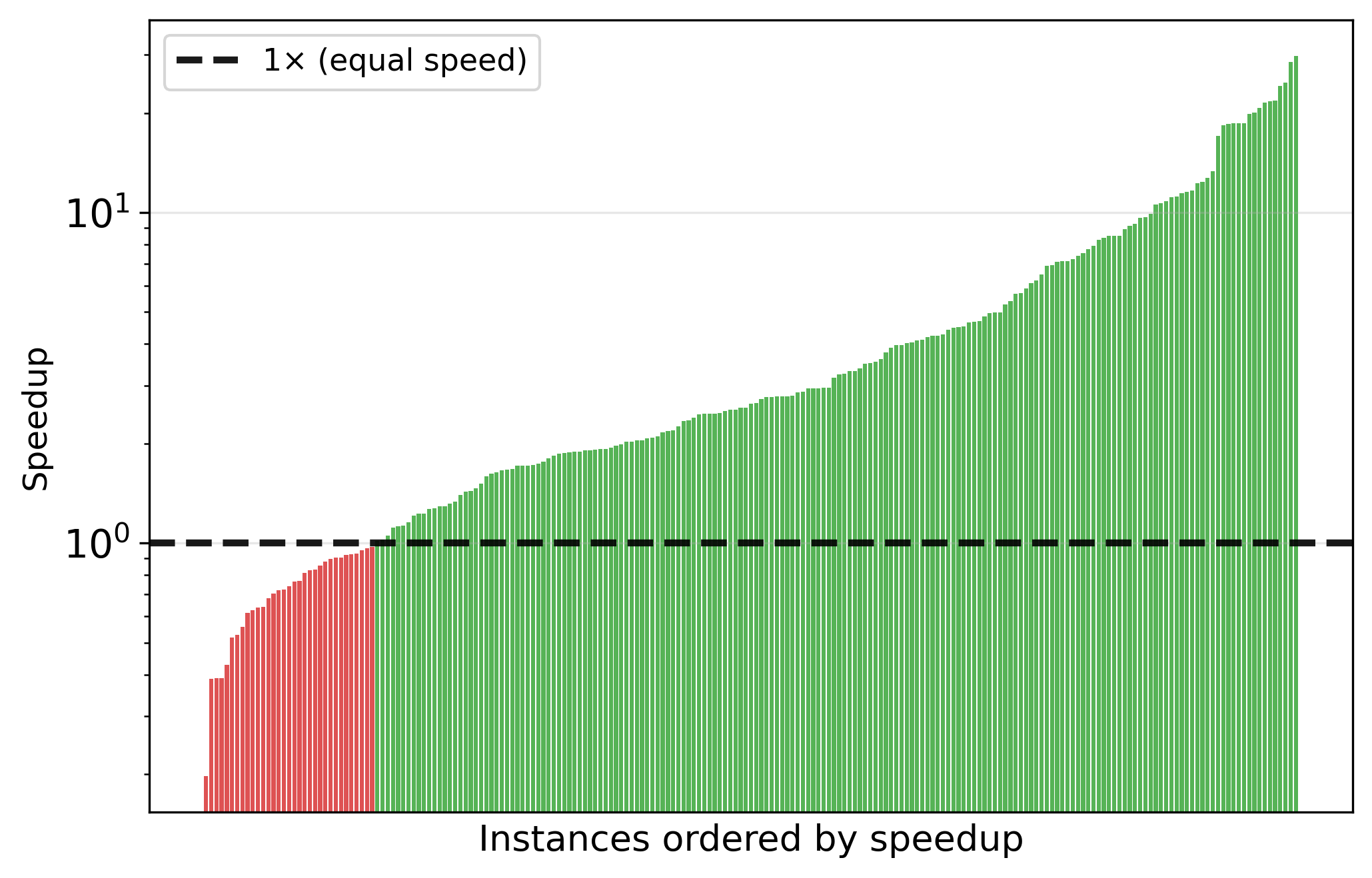}
    \caption{Improving the runtime of exact algorithm using $\Delta$Search for the MPS problem on North~\cite{di2000drawing} dataset.}
    \label{fig:exact}
    \vspace{-0.15in}
\end{figure}

Figure~\ref{fig:exact} shows the runtime of the modified RDS for the unweighted hereditary problem of Maximum Planar Subgraph on the North~\cite{di2000drawing} graph dataset. From the figure, we can see that in most graph instances, $\Delta$Search provides a massive speedup to the modified RDS, achieving a median speedup of $2.639{\times}$. The reason why $\Delta$Search slows down the search for some instances is that aggressive pruning might eliminate searches that could eventually yield a better bound. Although pruning will never prune an optimal solution, it might prune a solution that could serve as a better bound for future pruning. However, as can be seen from the figure, for most of the graph instances, $\Delta$Search accelerates the exact algorithm.

\section{Related Work}
\label{sec:relatedwork}

A significant amount of prior work has been done to unify optimization problem under general frameworks such as Constraint Programming (CP), Integer Linear Programming (ILP), generic branch and bound methods, meta-heuristic methods and, most recently, learning based approaches. Focus specifically on Graph Optimization Problems has been low but still existent such as the Hereditary Graph formulation and some learning based approaches. Most of these frameworks emphasize high generality and expressiveness, thus incurring significant scalability challenges or requiring problem-specific modeling and tuning. This gap motivates the need for a framework such as $\Delta$Search, whose formulation covers a large class of problems with similar underlying structure which also enables it to exploit the structure to enable efficient execution. $\Delta$Search also requires only a simpler underlying structure, making it more intuitive to formulate the graph problem in its domain.

\emph{Constraint Optimization Problems} (COP) provide a highly expressive framework in which problems can be encoded as variables with domains and constraints governing the feasible assignment of variables. COP can naturally capture a wide range of graph problems~\cite{russell2020modern} including selection problems (Maximum Independent Set, Maximum Vertex Cover, etc), labeling problems (Graph coloring), and assignment based problems such as Vehicle Routing Problem (VRP) and Facility location problems~\cite{brailsford1999constraint}. COP allows constraints to be stated in an intuitive and flexible manner. Advances in Constraint Programming (methodology for solving COP) has brought multiple strategies~\cite{dechter2003constraint} to improve performance such as look ahead strategies, domain filtering and clever backtracking. Despite these advances, Constraint Programming often suffers from poor scalability due to the large number of structural constraints inherent in large graphs which make constraint generation, propagation and repeated feasibility checks limit the performance. Practical success of COP in graph problems have usually been due to solver specific heuristic or extensive problem tuning. 

\emph{Integer Linear Programming} (ILP) and Mixed Integer Linear Programming (MILP) formulation allows graph problems to be encoded as linear constraints over integer variables. Although more restricted than COP, a large class of graph problems~\cite{almohamad2002linear, aneja1980integer, dias2025minimum} including minimum cut variants, facility location, routing, matching and partitioning problems have been successfully solved using ILP and MILP formulations. The success of ILP and MILP formulations is mainly due to the solvers which have been heavily researched and experimented. Translating graph problems into linear constraints and optimization functions introduces a significant abstraction burden which cannot be automated. This has resulted in no single general ILP/MILP framework for graph algorithms and the prior works have always manually abstracted the graph problems and then used the solvers. Translating graph problems into linear constraints also introduces a large number of variables and constraints as these formulations cannot naturally handle graph structures. These issues usually result in reduced performance from these solvers. When compared to COP, \cite{darby1998properties} noted that ILP/MILP formulations usually tend to do well when the search space is large with few constraints whereas COP formulations do well when the search space is highly constrained. In light of this, \cite{bockmayr1998branch, achterberg2008constraint, rodosek1999new} have tried combining both COP and MILP together to achieve better performance. 

\emph{Hereditary graphs properties} are those properties that are closed under vertex or edge deletions. Hereditary properties serve as a unification of multiple graph optimization problems where the goal is to find the maximum weighted subgraph. Many classical NP graph problems such as Maximum Clique, Maximum Independent Set Feedback Vertex Set problems fall into this category. Hereditary property problems are well-studied theoretically~\cite{bollobas1997hereditary, scheinerman1994size, borowiecki1997survey} and it also provides a more intuitive formulation than CSP or MILP formulations. While the underlying assumptions leads to more efficient exact algorithms~\cite{cohen2008generating, trukhanov2013algorithms} than naive search albeit still being exponential, it is highly restricted compared to CSP or ILP/MILP.

\emph{Branch and Bound frameworks} can broadly be applied to Graph optimization algorithms. Branch-and-bound (BnB)~\cite{galea2007bob++} systematically explores the solution space while pruning regions using upper and lower bounds on the objective. In principle, BnB can solve arbitrary combinatorial graph optimization problems. But the bounds always need to be problem-specific and even with the bounds, BnB frameworks tend to perform poorly without domain-tailored optimizations. \cite{morrison2016branch} presents a survey on how to implement BnB algorithms in multiple ways depending on the problem at hand. 

\emph{Meta-heuristics}~\cite{gonzalez2007handbook} such as Tabu Search, Greedy randomized adaptive search procedure (GRASP), Ant Colony Optimization, Evolutionary Algorithms, and Very-Large Scale Neighborhood Search provide general search templates that explore the solution space heuristically. These methods have been applied to a wide range of graph problems, including routing, clustering, partitioning, and scheduling. Meta-heuristics can scale to large instances and often produce high-quality solutions in practice despite the absence of optimality guarantees. But the scaling and performance are highly dependent on not just problem-specific tuning but also problem-specific implementation changes, which make it impossible for them to be treated as a black box for solving various graph problems. Metaheuristics also require heavy experimentation, as it is hard to know in advance which metaheuristic and which parameter configuration would work well for a problem~\cite{peres2021combinatorial}. They also cannot be applied indiscriminately to all problems as they require problem-specific implementation for most of the problems. Even a single metaheuristic will have variances in performance even for a single problem instance due to its stochastic nature.

\emph{Local Ratio}~\cite{bar2004local} is an approximation algorithm designed for optimization problems. Local Ratio works by breaking a given weight function $w$ into multiple simpler weight functions $w_1, w_2, \cdots , w_k$ where ${\Sigma_{i}} w_i = w$ and using a user-provided r-approximation algorithm for simpler weight functions to obtain an r-approximate solution for the general problem. Local ratio can be used to solve covering problems (e.g., Partial Vertex Cover, Feedback Vertex Set, Steiner Tree) and Packing problems (e.g., Independent Set, Interval Scheduling). Local Ratio achieves approximation guarantees, scalability and also applies to a large class of problems. But it is more complicated to use as the user has to provide a r-approximation algorithm for a simpler weighted instance. This puts the burden on the user to find simpler weight functions and algorithms to solve them.

\emph{Learning based approaches}~\cite{drori2020learning, peng2021graph, liu2022gon} have focused on creating unified architectures capable of solving multiple graph optimization problems by changing optimization function or training signal. These frameworks have been applied to graph optimization problems such as Minimum Spanning Tree, Traveling Salesman Problem, Vehicle Routing, Balanced Graph Partitioning and Maximum Independent Set. Unlike some of the previous formulations, learning-based approaches are graph-aware and can exploit the structural patterns across graph problems. But these approaches lack typical formal guarantees, are sensitive to training distribution and are often evaluated on a narrow set of problems and graph sizes and distributions. This makes their performance on out-of-distribution instances unpredictable. 

In contrast to the above approaches, the \emph{$\Delta$Search framework} targets a middle ground between expressiveness and efficiency while still providing an intuitive abstraction for any graph problem. Rather than relying on highly generic solvers or problem-specific algorithms, $\Delta$Search imposes a slightly restrictive assumption which still covers a large class of problems and which can be exploited to gain efficiency. This design allows $\Delta$Search to scale beyond traditional exact frameworks while retaining formal correctness guarantees, thereby addressing key limitations of existing unified graph optimization frameworks.

\vspace{-0.5em}
\section{Conclusions and Future Work}
\label{sec:conclusion}
In this paper, we introduced the Difference of Monotone formulation as a general reward-penalty decomposition that unifies Hereditary, Ancestral and other graph problems. We also presented $\Delta$Search, a heuristic that solves DoM problems with at most $O(n^2)$ calls to the scoring functions. We evaluated $\Delta$Search for a wide range of graph problems in which it surpassed the state of the art for UFLP and PCVC, remained competitive for MPS and achieved about 80-90\% of the solution quality for the remaining problems without any problem-specific tuning. We further demonstrated its effectiveness in pruning the search spaces for exact algorithms achieving around $2.6{\times}$ speedup for MPS. In future work, we plan to further improve $\Delta$Search, derive formal approximation guarantees and develop adaptive ordering strategies to improve solutions.

\vspace{-0.25em}
\section*{Data Availability Statement}
    The experimental graph datasets, code for $\Delta$Search and all baseline implementations used are available at \href{https://anonymous.4open.science/r/DeltaSearch-4F29/}{<Anonymous Github link>}.


\bibliographystyle{ACM-Reference-Format}
\bibliography{bibfile}

\end{document}